\documentclass[a4paper,12pt]{article}
\usepackage[english]{babel}
\usepackage[ansinew]{inputenc}
\usepackage{eurosym}
\usepackage{amsfonts}
\usepackage{mathrsfs}
\usepackage{amsthm}
\usepackage{amsmath, amssymb, bm}
\usepackage{mathtools}
\usepackage{graphics}
\graphicspath{{graphics/}}
\usepackage[usenames,dvipsnames]{xcolor}
\usepackage{epsfig}
\usepackage{rotating}
\usepackage{threeparttable}
\usepackage{booktabs}
\usepackage{multirow}
\usepackage{dcolumn}
\usepackage{hyperref}
\usepackage{apacite}
\usepackage[width=15cm,labelfont=bf]{caption}
\usepackage{pgfplots}
\usepackage{subcaption}

\usepackage{apacite}
\usepackage{graphicx}
\usepackage{setspace}
\usepackage{longtable}
\usepackage{colortbl}
\usepackage{xtab}
\usepackage{tikz,booktabs}
\usepackage[bottom]{footmisc}
\usepackage{indentfirst}
\usepackage{endnotes}
\usepackage{rotating}
\usepackage{lscape}
\usepackage{color, colortbl}%
\usepackage{authblk}
\usepackage{mdwlist}

\usepackage{verbatim}

\usepackage{caption}
\usepackage{longtable}

\pgfplotsset{width=10cm,compat=1.9}
\usepackage{calrsfs}
\DeclareMathAlphabet{\pazocal}{OMS}{zplm}{m}{n}

    \def\0{\mbox{\bf{0}}}

\providecommand{\U}[1]{\protect\rule{.1in}{.1in}}
\newtheorem{theorem}{Theorem}

\newtheorem{lemma}[theorem]{Lemma}

\newtheorem{proposition}[theorem]{Proposition}

\begin{document}

	\title{Testing the martingale difference hypothesis using martingale difference
		divergence function}
	 \author{ Luca Mattia Rolla\thanks{Department of Economics and Finance, Universita' degli Studi di Roma "Tor Vergata". Email: 
			lucamattia.rolla@alumni.uniroma2.eu}}
	\date{}
	\maketitle
	
	\begin{abstract}
		\noindent This article proposes a novel test for the martingale difference hypothesis based on the martingale difference divergence function, a recently developed dependence measure suitable for measuring the degree of conditional mean dependence of a random variable with respect to another. First, we discuss the use of martingale difference divergence in a time series framework as an alternative to the autocovariance function for detecting the existence of forms of nonlinear serial dependence. In particular, the measure equals zero if and only if the considered time-series components are conditionally mean-independent. This characteristic makes it suitable for studying the behavior of white noise processes characterized by non-null mean conditional on the past. We discuss the asymptotic properties of sample martingale difference divergence in a univariate time series framework, refining some of the results existing in the literature. Doing this allows us to build a Ljung-Box-type test statistic by summing the sample martingale difference divergence function over a finite number of lags. Under suitable conditions, the asymptotic null distribution of our test statistic is also established. The finite sample performance is discussed via a Monte Carlo study as we demonstrate its consistency against uncorrelated non-martingale processes. Finally, we show an empirical application for our methodology in analyzing the properties of the Standard and Poor's 500 stock index.
		
	\end{abstract}
	
	 \par\noindent\textit{Keywords}: Conditional mean independence, Testing, Nonlinearity, Martingale difference hypothesis.

	\clearpage
	
	\section{\ Introduction}
	\label{sec:intro}
	\noindent The martingale difference hypothesis (MDH) underlies many theories in
	economics and finance by defining a more stringent condition than
	uncorrelatedness. In general terms, the future values of a martingale difference sequence (MDS) are unpredictable
	using current information so that the best predictor in the mean squared
	error sense is simply the unconditional expectation. In this paper we consider testing the hypothesis that, given a real-valued stationary time series $\left\{  X_{t}\right\}
	$, we have%
	\begin{equation}
		E\left(  X_{t}|X_{t-1},X_{t-2},...\right)  =E\left(  X_{t}\right), \text{         \\        \\    \\  a.s.}.
		\label{eq:First}
	\end{equation}

The MDH mandatorily implies the absence of any linear serial dependence regarding serial dependence structures. The converse is not valid, as a time series can have zero autocorrelation while exhibiting some form of nonlinear serial dependence in the conditional mean. In general, it is not sufficient to test for the MDH by simply considering the autocorrelation function, and we need to distinguish between an MDS and a white noise process by accounting for the existence of nonlinear dependence in the mean. In this first section, we discuss the problem of accounting for the various facets of nonlinearity in time series and introduce some measures of nonlinear dependence suited for dealing with it. As we shall see, it is possible to employ their respective properties to test a variety of serial dependence hypotheses, including the one of MDH.

	\subsection*{Nonlinearity in general}
	The analysis of time series temporal dependence has traditionally focused on the analysis of their autocorrelation function or, equivalently, of their spectral density. These tools, however, become inappropriate in the analysis of purely nonlinear and non-Gaussian stochastic processes. To address these limitations, \cite{hong1999hypothesis} introduced the so-called generalized spectral density function, which is explicitly
	devised to take nonlinearities into consideration and to put to the test a wide range of dependence hypotheses in a nonlinear time series framework. Let $\left\{  X_{t},\text{ }t\in\mathbb{Z}\right\}  $ be a univariate, strictly stationary, time
	series having joint characteristic function $\phi\left(  u,v\right)  =E\left(
	e^{i\left(  uX_{t}+vX_{t-j}\right)  }\right)  $ and marginal characteristic
	functions $\phi\left(  u\right)  =E\left(  e^{iuX_{t}}\right)  $ and
	$\phi\left(  u\right)  =E\left(  e^{ivX}\right)  $ defined for the pair
	$\left(  X_{t},X_{t-\left\vert j\right\vert }\right)  $, for $\left(
	u,v\right)  \in\mathbb{R}^{2}$ and $i^{2}=-1$. Hong's methodology is based on the analysis of the
	Fourier transform of the generalized covariance function defined as%
	
	\begin{align}
		\sigma_{j}\left(  u,v\right)   &  =Cov\left(  e^{iuX_{t}},e^{ivX_{t-j}}\right)
		\\
		&  =\phi\left(  u,v\right)  -\phi\left(  u\right)  \phi\left(  v\right).
		\nonumber
	\end{align} It is clear that, for
	any given pair $\left(  u,v\right)  ,$ the function $\sigma_{j}\left(  u,v\right)
	$ equals zero if and only if $\left\{  X_{t}\right\}  $ is serially
	independent at lag $j$, i.e., such a function makes it possible  to capture the presence of any form pairwise dependence, regardless of whether the series shows signs of linear serial dependence or not. In particular, we consider the following regularized norm 
	\begin{align}
		\label{eq:normTermSigmauv}\left\vert \left\vert \sigma_{j}\left(  u,v\right)
		\right\vert \right\vert _{W}^{2}  &  =\left\vert \left\vert Cov\left(
		e^{iuX_{t}},e^{ivX_{t-\left\vert j\right\vert }}\right)  \right\vert
		\right\vert ^{2}  \\ 
		&  =\int_{\mathbb{R}^{2}}\left\vert \sigma_{j}\left(  u,v\right)  \right\vert
		^{2}dW\left(  u,v\right)  ,\nonumber
	\end{align} where such a norm term involves integrating defined over the space of the
	parameters in the characteristic functions, with $W\left(  u,v\right)  $ being an arbitrary weight function for which the integral is finite. In its base form, the theory of generalized spectral density can be used to define a test statistic for the null hypothesis of serial independence, this one being described in terms of the norm of the generalized covariance sample estimator $ \widehat{\sigma}_{j}\left(
		u,v\right)$.\footnote{Estimation of $\sigma_{j}\left(  u,v\right)$ involves estimating the series' joint and marginal characteristic functions. In this sense, the concept of empirical characteristic function will be introduced in Section \ref{sec:spectralD}}

 \noindent Based on Hong's work, a novel test for the hypothesis of
	serial independence has been introduced by \cite{fokianos2017consistent}. The authors provide a closed form expression for the norm component in Equation (\ref{eq:normTermSigmauv}) in terms of distance covariance function (\cite{szekely2007measuring}). First defined for the i.i.d. case, distance covariance allows identifying any nonlinear dependence pattern that the covariance function would neglect. In its standardized form, named distance correlation, it can be interpreted as a generalization of classical Pearson correlation to a framework including possibly nonlinear relationships. \cite{zhou2012measuring}extended the use of the measure to time series by introducing the so-called auto-distance covariance (correlation) function, a generalization of classical autocovariance capable of identifying any serial link between any time series component and its past values. More recently \cite{fokianos2017consistent} show how auto
	distance covariance function gives us a way to explicitate the 
	norm in Equation (\ref{eq:normTermSigmauv}) under some mild assumptions about the weight functions used to compute the integral.  Doing this greatly simplifies the
	expression of Hong's test statistic, while keeping all the beneficial characteristic from the original formulation.

 \noindent The test statistic defined in Section (\ref{sec:martTest}) has theoretical foundations in the above-described approach. It is, however, built around a different definition of the generalized covariance term that allows us to focus on the MDH rather than on the broader hypothesis of serial independence.

	\subsection*{Testing the Martingale Difference Hypothesis}
		
	\noindent In the literature, several works have addressed the problem of testing the MDH. Many of the proposed test statistics, however, due to their reliance on forms of dependence found in the second moments only, could lead to ignoring various forms of nonlinear dependence characterizing the conditional mean, in this way misinterpreting the nature of the studied processes. In particular, these tests suffer from the severe limitation of being inconsistent against non-martingale difference sequences with zero autocorrelations, that is, when only nonlinear dependence is present.\footnote{Consider for instance \cite{durlauf1991spectral} and \cite{deo2000spectral}.} This family of processes
	includes, for instance, the Bilinear (BIL), the Threshold autoregressive (TAR)
	and the Nonlinear moving average (NLMA) processes.
 
 \noindent Various authors have developed methodologies to address this issue:  \cite{hong1999hypothesis} was among the first ones to implement a test for MDH that allowed to cope with the nonlinear features characterizing the conditional mean. Indeed, the generalized spectral density approach offers a flexible framework for testing several different and specific hypotheses of serial dependence. As discussed in Section (\ref{sec:spectralD}), the generalized spectral density function can be differentiated at the origin with respect to the auxiliary parameters in the characteristic functions to obtain different specifications of the norm term in Equation (\ref{eq:normTermSigmauv}). By computing partial derivatives of various orders, it is possible to obtain several different test specifications to examine numerous aspects of dependence.

 \noindent More importantly, for the sake of our study, the baseline function can be appropriately differentiated to construct a test specification suitable for dealing with the	MDH. In the case of our interest, the generalized spectral derivative is an expression of the following generalized covariance term
	\begin{align}
		\sigma_{j}\left(  v\right)   &  =Cov\left(  X_{t},e^{ivX_{t-\left\vert
				j\right\vert }}\right) \\
		&  =E\left(  X_{t}e^{ivX_{t-\left\vert j\right\vert }}\right)  -E\left(
		X_{t}\right)  \phi\left(  v\right)  ,\nonumber
	\end{align}
	with $\phi\left(  v\right)  $, $v\in\mathbb{R}$, being the characteristic function of
	the variable $X_{t-j}$.

 \noindent Following \cite{hong1999hypothesis}, \cite{hong2003inference} and \cite{escanciano2006generalized}, our approach for testing the MDH will be, in particular, to consider the hypothesis
	\begin{equation}
		H_{0}:E\left[  X_{t}-E\left(  X_{t}\right)  |X_{t-j}\right]  =0, \text{\ \ } \forall j\geqslant1, \text{\ \ } a.s.,
		\label{eq:null}
	\end{equation}
where, in particular, we consider all the pairwise implications  In particular, as discussed in \cite{bierens1982consistent}, it is possible to express the null hypothesis in Equation (\ref{eq:null}) in the following terms
\begin{equation}
	H_{0}\Longleftrightarrow \sigma _{j}\left( v\right) =0\text{ \ \ \ }\forall
	j\geqslant 1,\text{ \ almost everywhere in $v$\ } 
\end{equation}
	Accordingly, the test for the MDH introduced by Hong (1999) is defined in terms of the regularized norm%
	\begin{align}
		\label{eq:alfio}\left\vert \left\vert \sigma_{j}\left(  v\right)  \right\vert
		\right\vert _{w}^{2}  &  =\left\vert \left\vert Cov\left(  X_{t}%
		,e^{ivX_{t-\left\vert j\right\vert }}\right)  \right\vert \right\vert^{2} \\
		&  =\int_{\mathbb{R}}\left\vert \sigma_{j}\left(  v\right)  \right\vert
		^{2}dW\left(  v\right)  ,\nonumber
	\end{align}
	for a specified weighting function $W\left(  v\right)  $. 
 
 \noindent Following an approach similar to that of \cite{fokianos2017consistent}, we can simplify the original test formulation by providing a closed-form expression for the norm in Equation (\ref{eq:alfio}) thanks to the recent introduction of martingale difference divergence (\cite{shao2014martingale}) as a measure of conditional mean dependence between random variables. The martingale difference divergence function belongs to the same family of dependence measures as distance covariance, and, like the latter, it involves some specific assumptions about the formulation of the weight function used in the norm of the generalized covariance term. The testing procedure introduced in this paper has to be seen as complementary to the one described in \cite{fokianos2017consistent}: the two tests can be jointly used to get a deeper understanding of the nature of the series and to focus on different aspects of dependence (linear or nonlinear, in the conditional mean or not). From a purely computational standpoint, the novel specification for these serial dependence tests allows us to avoid any form of numerical integration that is instead required in the original definition of the statistics.
	
	Our statistic can be compared to the tests for the MDH by \cite{escanciano2006generalized} and \cite{wang2022testing}, which appear to have similar formulations and properties. The first one also has its foundation in Hong's approach and is based on the related concept
	of the generalized spectral distribution function. In this case, the authors obtain an explicit formulation
	for the norm term from Equation (\ref{eq:alfio}) by assuming that the weighting
	function in the integral corresponds to a proper probability distribution function (for instance, the cdf of standard normal). Regarding the test by \cite{wang2022testing}, its definition is based on a multivariate extension of martingale difference divergence, martingale difference divergence matrix (\cite{lee2018martingale}). Its similarity with our statistic is remarkable, even though it originates from an essentially different theoretical construction. Using the martingale difference divergence matrix allows the authors to obtain a test for the MDH that applies to the study of vector time series. Even so, in this paper, we shall focus on testing for the dependence properties of univariate time series. We will discuss the main features of the competing test specifications in Section (\ref{section:empirics}) when we compare their size and power properties to that of our proposed statistic. 
 
\begin{figure}[ptb]
		\centering
		\begin{subfigure}[b]{\textwidth}
			\centering
			\includegraphics[height=0.48\textwidth,width=\textwidth]{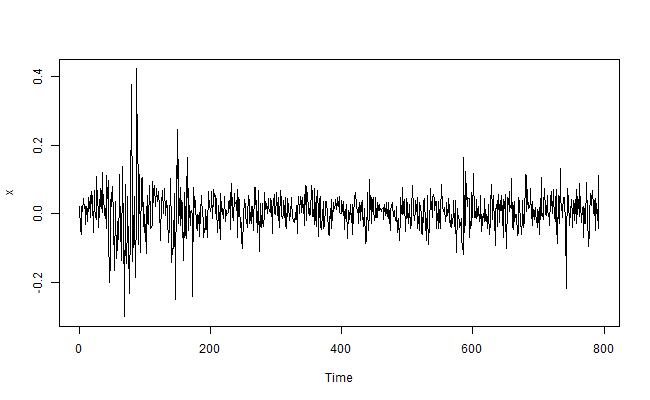}
		\end{subfigure}
		\par
		\centering
		\begin{subfigure}[b]{0.49\textwidth}
			\centering
			\includegraphics[width=\textwidth]{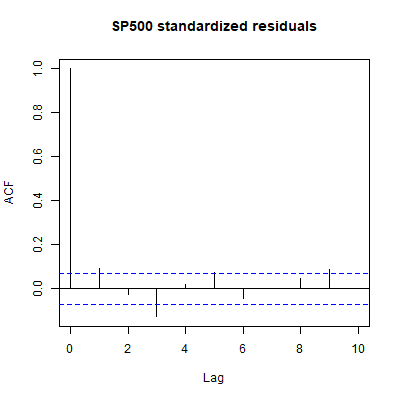}
		\end{subfigure}
		\begin{subfigure}[b]{0.49\textwidth}
			\centering
			\includegraphics[width=\textwidth]{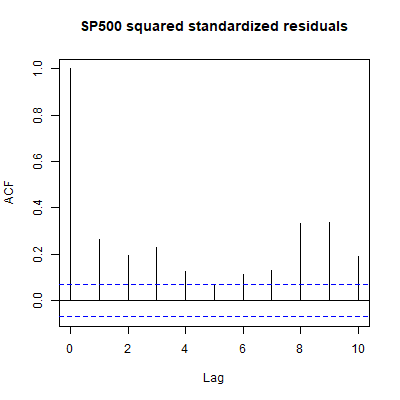}
		\end{subfigure}
		\par
		\centering
		\begin{subfigure}[b]{0.49\textwidth}
			\centering
			\includegraphics[width=\textwidth]{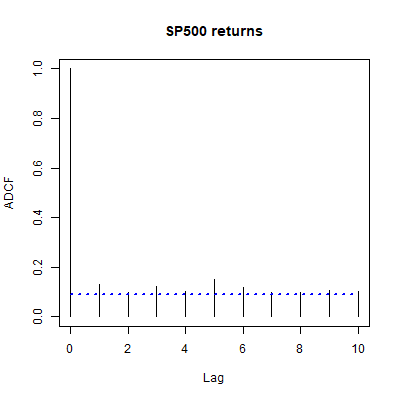}
		\end{subfigure}
		\begin{subfigure}[b]{0.49\textwidth}
			\centering
			\includegraphics[width=\textwidth]{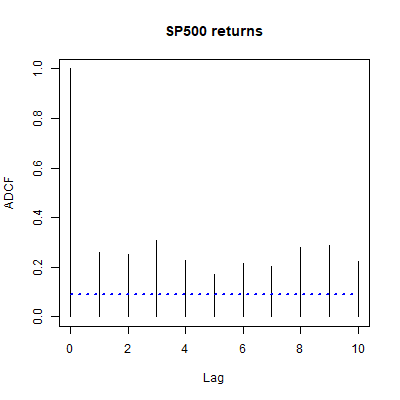}
		\end{subfigure}
		\caption{ S\&P 500 excess returns series (upper panel), ACF plot (middle
			panel) and ADCF plot for returns and squared returns (lower panel).}%
		\label{spSEries}%
	\end{figure}

 To finally conclude this introduction to the contents of the paper, Section (\ref{spSEries}) illustrates a brief empirical application of our methodology to the monthly excess returns
series of the S\&P 500 index.
	
	\section{Generalized Spectral Density and its Derivatives}
	\label{sec:spectralD}
	We start our analysis by introducing \cite{hong1999hypothesis}'s generalized spectral density
	theory. Given a strictly stationary, strong mixing process $\left\{
	X_{t}\right\}  $, we assume it has marginal characteristic function
	$\phi\left(  u\right)  =Ee^{iuX_{t}}$ and a pairwise joint characteristic
	function $\phi_{j}\left(  u,v\right)  =Ee^{i\left(  uX_{t}+vX_{t-\left\vert
			j\right\vert }\right)  }$, where $i=\sqrt{-1},u,v\in\left(  -\infty
	,\infty\right)  $ and $j=0,\pm1,...$. As already mentioned, the generalized
	covariance computed for the transformed variables $e^{i\left(  uX_{t}\right)
	}$ and $e^{i\left(  vX_{t-\left\vert j\right\vert }\right)  }$ is defined as
	$\sigma_{j}\left(  u,v\right)  =Cov\left(  e^{i\left(  uX_{t}\right)
	},e^{i\left(  vX_{t-\left\vert j\right\vert }\right)  }\right)  $, where it
	possible to write, equivalently,$\ \sigma_{j}\left(  u,v\right)  =\phi
	_{j}\left(  u,v\right)  -\phi\left(  u\right)  \phi\left(  v\right)  .$ Since
	$\phi_{j}\left(  u,v\right)  =\phi\left(  u\right)  \phi\left(  v\right)  $
	for all $u,v$ if and only if $\left\{  X_{t}\right\}  $ and $\left\{
	X_{t-\left\vert j\right\vert }\right\}  $ are independent, the covariance term
	$\sigma_{j}\left(  u,v\right)  $ equals zero if and only if the variable in
	consideration is independent of its past (at lag $j$) value. Hong is able to
	exploit this feature $\sigma_{j}\left(  u,v\right)  $ by studying its Fourier
	transform: assuming that $\sup_{u,v\in\left(  -\infty,\infty\right)  }%
	\Sigma_{j}\left\vert \sigma_{j}\left(  u,v\right)  \right\vert <\infty,$ the
	Fourier transform of $\sigma_{j}\left(  u,v\right)  $ exists and is defined
	as\smallskip%
	\begin{equation}
		f\left(  u,v,\omega\right)  =\frac{1}{2\pi}%
		{\displaystyle\sum_{j=-\infty}^{\infty}}
		\sigma_{j}\left(  u,v\right)  e^{-ij\omega},\text{ \ \ }\omega\in\left[
		-\pi,\pi\right]  .
	\end{equation}
	\smallskip and is named generalized spectral density.\newline The generalized
	spectral density $f\left(  u,v,\omega\right)  $ can capture all pairwise
	dependencies in contrast to standardized spectral density, which, on the other hand, can determine only linear dependencies. Hong shows that $f\left(
	u,v,\omega\right)  $ can be consistently estimated by \smallskip%
	\begin{equation}
		\widehat{f}_{n}\left(  \omega,u,v\right)  =\frac{1}{2\pi}%
		{\displaystyle\sum_{j=-\left(  n-1\right)  }^{n-1}}
		\left(  1-\frac{\left\vert j\right\vert }{n}\right)  ^{1/2}k\left(
		j/p\right)  \widehat{\sigma}_{j}\left(  u,v\right)  e^{-ij\omega}%
	\end{equation}
	\smallskip where $\widehat{\sigma}_{j}\left(  u,v\right)  =\widehat{\phi}%
	_{j}\left(  u,v\right)  -\widehat{\phi}_{j}\left(  u\right)  \widehat{\phi
	}_{j}\left(  v\right)  $ and $\widehat{\phi}_{j}\left(  u,v\right)  =\left(
	n-\left\vert j\right\vert \right)  ^{-1}\sum_{t=\left\vert j\right\vert
		+1}^{n}e^{i\left(  uX_{t}+vX_{t-\left\vert j\right\vert }\right)  }$ is the
	empirical joint characteristic function. Finally, the variable $p=p_{n}$ is a
	bandwidth or lag order parameter and $k\left(  \cdot\right)  $ is a symmetric
	kernel function or "lag window". Under the assumption of serial independence,
	$f\left(  \omega,u,v\right)  $ becomes the flat spectrum%
	\[
	f_{0}\left(  \omega,u,v\right)  =\frac{1}{2\pi}\sigma_{0}\left(  u,v\right)
	,\text{ \ \ }\omega\in\left[  -\pi,\pi\right]  ,
	\]
	which can be equivalently estimated by $\widehat{f}_{0}\left(  \omega
	,u,v\right)  =\left(  2\pi\right)  ^{-1}\widehat{\sigma}_{0}\left(
	u,v\right)  .$\newline In order to detect serial dependence, we can compare the
	two estimators $\widehat{f}_{n}\left(  u,v,\omega\right)  $ and $\widehat
	{f}_{0}\left(  u,v,\omega\right)  $: in \cite{hong1999hypothesis} this is done by means of a
	weighted $L_{2}$ distance
	\begin{align}
		H_{99}  &  =\int\int_{-\pi}^{\pi}\left\vert \widehat{f}_{n}\left(
		\omega,u,v\right)  -\widehat{f}_{0}\left(  \omega,u,v\right)  \right\vert
		^{2}d\omega dW\left(  u,v\right)  		\label{eq:hongIndtest}  \\
		&  =\left(  \frac{\pi}{2}\right)  ^{-1}\sum_{j=1}^{n-1}k^{2}\left(
		j/p\right)  \left(  1-j/n\right)  \int\left[  \widehat{\sigma}_{j}\left(
		u,v\right)  ^{2}\right]  dW\left(  u,v\right) \nonumber\\
		&  =\left(  \frac{\pi}{2}\right)  ^{-1}\sum_{j=1}^{n-1}k^{2}\left(
		j/p\right)  \left(  1-j/n\right)  \left\vert \left\vert \widehat{\sigma}_{j}\left(
		u,v\right)  \right\vert \right\vert _{W}^{2}\nonumber
	\end{align}
	where the second equality follows from Parseval's identity and $W\left(
	u,v\right)  $ is a specified symmetric weighting function (as specified in
	\cite{hong1999hypothesis}, one possible choice for the weighting function is represented by
	the standard normal CDF).\newline The statistic $H_{99}$ is used by Hong to
	put to test the hypothesis that a given series is an i.i.d. sequence: still,
	the generalized spectral density framework offers enough flexibility to easily
	put to test a great number of different aspects related to serial dependence.

	\subsection*{Using Distance Covariance to test for serial independence}
	
	As discussed already in Section (\ref{sec:intro}), the auto-distance covariance (ADCV) function gives us a way to express the norm
	term in Equation (\ref{eq:hongIndtest}), in this way simplifying Hong's original expression. In particular, the ADCV function at lag $j\in\mathbb{Z}$ is defined as
	the non-negative square root of the $L_{2}$ norm
	\begin{align}
		V_{X}^{2}\left(  j\right)   &  = \text{\ \ } \left\vert \left\vert \sigma_{j}\left(  u,v\right)  \right\vert
		\right\vert _{\mathcal{W}}^{2}  \\ 
		&  =  \int_{\mathbb{R}^{2}}\left\vert
		\sigma_{j}\left(  u,v\right)  \right\vert ^{2}\mathcal{W}\left(
		u,v\right)  dudv \nonumber\\
		&  = \frac{1}{\pi^{2}}\int_{\mathbb{R}^{2}}\frac{\left\vert \sigma 
			_{j}\left( u,v\right)  \right\vert ^{2}}{\left\vert u%
			\right\vert ^{2}\left\vert v\right\vert ^{2}}du%
		dv \text{, \ \ }j=0,\pm1,\pm2,...\nonumber %
	\end{align}

where we assume that the weight function $W\left(  u,v\right)  $ exactly corresponds to that defined in \cite{szekely2007measuring}, which, in the univariate time series case equals \begin{equation}
    W\left(  u,v\right)  =\mathcal{W}\left(  u,v\right)
	=\left(  \pi^{2}\left\vert u\right\vert ^{2}\left\vert v\right\vert
	^{2}\right)  ^{-1}.
\end{equation} Given the fact that the joint characteristic function factorizes into the product of its marginals under independence, $V_{X}^{2}$ equals 0 if and only if $X_t$ and $X_{t-\vert j \vert}$ are independent. The standardized form of ADCV, named auto-distance correlation function (ADCF), takes the form

\begin{equation}
R_{X}^{2}\left(  j\right)  =\left\{
\begin{array}
[c]{c}%
\frac{V_{X}^{2}\left(  j\right)  }{V_{X}^{2}\left(
0\right)  },\text{ \ }V_{X}^{2}\left(  0\right)  >0;\\
0,\text{ \ otherwise.}%
\end{array}
\right.
\end{equation}
 is bounded in the unit interval and equals zero under independence. In this sense,
ADCF can be seen as a generalization of classical ACF, capturing all pairwise dependencies, including those with zero autocorrelation. For this reason, the measure
is suitable for exploring possible non-linear dependence relationships in time series that are not detected by the ACF.

Let us define $Y_{t}=X_{t-\vert j \vert}$. Based on the sample $\{ \left( X_t,Y_t\right):t=1+ \vert j \vert ,..., n \}$, the sample ADCV is defined from the $\left(  n-\vert j \vert\right)
\times\left(  n-\vert j \vert\right)  $ pairwise Euclidean distance matrices $a$ and $b$ of elements
$ a_{rl}  =\left\vert X_{r}-X_{l}\right\vert $
and $  b_{rl}  = \left\vert Y_{r}-Y%
_{l}\right\vert $. These matrices are double-centered, by row and column means, so that we consider the matrices $A$, $B$ of elements
\begin{align}
A_{rl}=a_{rl}-\overline{a}_{r.}-\overline{a}%
_{.l}+\overline{a}_{..},\text{ \ \   }
  B_{rl}=b_{rl}-\overline{b}_{r.}-\overline
{b}_{.l}+\overline{b}_{..}.
\end{align}
where $\overline{a}_{r.}= \sum_{l=1}^{n}a_{rl}  /n, \text{ \   } 
 \overline{a}_{.l}=  \sum_{l=r}^{n}a_{rl}  /n, \text{ \  } \overline
 {a}_{..}=  \sum_{r,l=1}^{n}a_{rl}  /n^{2}$ and analogously for the terms
 $\overline{b}_{r.}$, $\overline{b}_{.l}$. Accordingly, sample ADCV is defined as the nonnegative square
root of
\begin{equation}
	\widehat{V}_{X}^{2}\left(  j\right)  =\frac{1}{\left(  n-\vert j \vert\right)
		^{2}}\sum_{r,l=1}^{n-j}A_{rl}B_{rl}%
\end{equation}
where $\widehat{V}_{X}^{2}\left(  j\right) \geq 0$ a.s., for $j \in \mathbb{Z}$, where the equality holds under independence. In the univariate case, under the assumption that $X_{t}$ is a strictly
stationary $\alpha$-mixing process and $E\left(  X_{t}\right)  <\infty$,
\cite{fokianos2017consistent} \ proved the convergence almost surely of
$\widehat{V}_{X}^{2}\left(  \cdot\right)  $ to its population counterpart for
a fixed lag value $j$. The test for the hypothesis of serial independence
	introduced by \cite{fokianos2017consistent} rexpresses the statistic in Equation (\ref{eq:hongIndtest}) by using the ADCV function. Accordingly, the test statistic takes now the form
	\begin{equation}
	T_{n}=\sum_{j=1}^{n-1}\left(  n-j\right)  k^{2}\left(  j/p\right)  \widehat
	{V}_{X}^{2}\left(  j\right)  ,
	\end{equation}
	where this simplified expression is computationally more efficient by avoiding the computation of any integral. The construction of our test for the MDH
	based on martingale difference divergence takes inspiration from this approach.
	
	\section{Testing the Martingale Difference Hypothesis}
		\label{sec:martTest}

 While the generalized spectral density in its base form
	is used to capture any form of serial dependence, its partial derivatives (at zero) can be used to test for an array of more specific hypotheses as for
	example zero serial correlation, the MDH, conditional homoscedasticity,
	conditional symmetry and many others. For this purpose, we can consider the $L_{2}$ distance between the  generalized spectral density derivatives estimators%
	\begin{gather}
		\widehat{f}_{n}^{\left(  0,m,l\right)  }\left(  \omega,u,v\right)  =\frac
		{1}{2\pi}%
		{\displaystyle\sum_{j=-\left(  n-1\right)  }^{n-1}}
		\left(  1-\left\vert j\right\vert /n\right)  ^{1/2}k\left(  j/p\right)
		\widehat{\sigma}_{j}^{\left(  m,l\right)  }\left(  u,v\right)  e^{-ij\omega}\\
		\widehat{f}_{n}^{\left(  0,m,l\right)  }\left(  \omega,u,v\right)  =\frac
		{1}{2\pi}\widehat{\sigma}_{0}^{\left(  m,l\right)  }\left(  u,v\right)
		\nonumber
	\end{gather}
	\smallskip where $\widehat{\sigma}_{j}^{\left(  m,l\right)  }\left(
	u,v\right)  =\partial^{m+l}\widehat{\sigma}_{j}\left(  u,v\right)
	/\partial^{m}u\partial^{l}v.$ Specific aspects of dependence can be considered
	depending on the order $\left(  m,l\right)  :$ this peculiar property derives
	from the various specifications that the generalized covariance term assumes according to the order of differentiation. In particular, when the
	order of differentiation is taken to be $\left(  m,l\right)  =\left(
	1,0\right)  $, the sample estimator of the spectral density derivatives can be used to construct a test for the MDH as 
	\begin{align}
		&  \int\int_{-\pi}^{\pi}\left\vert \widehat{f}_{n}^{\left(  1,0\right)
		}\left(  \omega,0,v\right)  -\widehat{f}_{0}^{\left(  1,0\right)  }\left(
		\omega,0,v\right)  \right\vert ^{2}d\omega dW\left(  v\right) 
		\label{eq:hongtest}		\\
		&  =\left(  \frac{\pi}{2}\right)  ^{-1}\sum_{j=1}^{n-1}k^{2}\left(
		j/p\right)  \left(  1-j/n\right)  \int\left\vert \widehat{\sigma}_{j}^{\left(
			1,0\right)  }\left(  0,v\right)  \right\vert ^{2}dW\left(  v\right)
		\nonumber\\
		&  =\left(  \frac{\pi}{2}\right)  ^{-1}\sum_{j=1}^{n-1}k^{2}\left(
		j/p\right)  \left(  1-j/n\right)  \left\vert \left\vert \widehat{\sigma}%
		_{j}^{\left(  1,0\right)  }\left(  v\right)  \right\vert \right\vert _{W}%
		^{2},\nonumber
	\end{align}
	where the second equality follows from Parseval's identity. The estimator
	\begin{equation}
	    \widehat{\sigma}_{j}^{\left(  1,0\right)  }\left(  0,v\right)  =\left(
	n-j\right)  ^{-1}%
	{\displaystyle\sum_{t=j+1}^{n}}
	X_{t}\left(  e^{ivX_{t-j}}-\widehat{\phi}_{j}\left(  0,v\right)  \right)  \end{equation} is
	consistent for $\sigma_{j}^{\left(  1,0\right)  }\left(  0,v\right)
	=Cov\left(  X_{t},e^{ivX_{t-\left\vert j\right\vert }}\right)  =E\left(
	X_{t}e^{ivX_{t-\left\vert j\right\vert }}\right)  -E\left(  X_{t}\right)
	\phi\left(  v\right)  $. Recall that the term $\sigma
	_{j}^{\left(  1,0\right)  }\left(  0,v\right)  $ has been  introduced in Section \ref{sec:intro}
	as $\sigma_{j}\left(  v\right)  $. For the rest of analysis we shall therefore adopt this simplified notation. 
	
	\subsection*{Testing the Martingale Difference Hypothesis \newline using
		Martingale difference divergence function}

 As an extension of distance covariance, \cite{shao2014martingale} introduced the
so-called martingale difference divergence and its standardized analog,
martingale difference correlation, as a measure of the conditional mean (in)dependence between two variables. The concept was later introduced into a time series framework by \cite{lee2018martingale}, who discussed the matrix extension of the measure termed martingale difference divergence matrix. Similarly to the auto-distance covariance function, we can define the time series analog
for martingale difference divergence, which we refer to as
the auto-martingale difference divergence function (AMDD). As seen already for ADCV, AMDD gives us a simplified way to express the
	norm term $\left\vert \left\vert \sigma_{j}\left(  v\right)  \right\vert
	\right\vert _{W}$ by assuming that the weight function under the integral
	 takes the form specified by \cite{szekely2007measuring}. 
In particular we define AMDD at lag $j\in\mathbb{Z}$ defined as the nonnegative square root of
\begin{align}
		MDD_{X}^{2}\left(  j\right)   &  =\left\vert \left\vert \sigma_{j}\left(
		v\right)  \right\vert \right\vert _{\mathcal{W}}^{2}\\
		&  =\int_{\mathbb{R}}\left\vert \sigma_{j}\left(  v\right)  \right\vert
		^{2}d\mathcal{W}\left(  v\right)  \nonumber
		\\
		&  =\frac{1}{\pi}\int_{\mathbb{R}}\frac{\left\vert \sigma_{j}\left(
			v\right)  \right\vert ^{2}}{\left\vert v\right\vert ^{2} %
		}dv \text{, \ \ }j=0,\pm1,\pm2,...\nonumber
	\end{align}
\noindent where the nonintegrable weight function ${\mathcal{W}} \left(  \cdot\right)  $ 
 simplifies in
this case to ${\mathcal{W}} \left(  v\right)  =\left(  \pi \left\vert
v\right\vert ^{2}\right)  ^{-1}$. Such a function will equal zero almost surely if and only if for a fixed
$j\in\mathbb{N}^{+}$ the condition $E\left(  X_{t}%
|X_{t-j}\right)  =E\left(  X_{t}\right)  $ is satisfied. The standardized form of AMDD, named auto-martingale difference correlation function (AMDCF)
is defined as the nonnegative square root of

\begin{equation}
	MDC_{X}^{2}\left(  j\right)  =\left\{
	\begin{array}
		[c]{c}%
		\frac{MDD_{X}^{2}\left(  j\right)  }{\sqrt{Var^2\left(  X_{t}\right)
				V^{2}\left(  0\right)  }},\text{ \ }Var^2\left(  X%
		_{t}\right)  V_{x}^{2}\left(  0\right)  \neq0;\\
		0,\text{ \ \ \ \ \ \ \ \ \ \ otherwise}
	\end{array}
	\right.
\end{equation}
taking values in the unit interval. It is clear how AMDC represents a more narrow measure of dependence compared to ADCF, allowing us to focus on a specific aspect of dependence, such as conditional mean dependence. 
	
Let us define $Y_{t}=X_{t-\vert j \vert}$. Based on the sample $\{ \left( X_t,Y_t\right):t=1+ \vert j \vert ,..., n \}$, the sample AMDD function is defined from the $\left(  n-\vert j \vert\right)
\times\left(  n-\vert j \vert\right)  $ pairwise Euclidean distance matrices $a$ and $b$ of elements
$ a_{rl}  =\left\vert X_{r}-X_{l}\right\vert $
and $  b_{rl}  = \frac{1}{2} \left\vert Y_{r}-Y%
_{r}\right\vert^{2} $, where the variable $Y_{t}$ is simply defined as
$Y_{t}=X_{t-j}.$ The statistic is the non-negative square
root of
	\begin{equation}
	\widehat{MDD}_{X}^{2}\left(  j\right)  =\left(  n-j
	\right)  ^{-2}\sum_{r,l=1+\left\vert j\right\vert }^{n}A_{rl}B_{rl}%
\end{equation}
defined for the two double-centered  matrices $A$ and $B$ with elements
\begin{align}
A_{rl}=a_{rl}-\overline{a}_{r.}-\overline{a}%
_{.l}+\overline{a}_{..}, \text{ \ \   }
  B_{rl}=b_{rl}-\overline{b}_{r.}-\overline
{b}_{.l}+\overline{b}_{..}.
\end{align}
where $\overline{a}_{r.}= \sum_{l=1}^{n}a_{rl}  /n, \text{ \   } 
 \overline{a}_{.l}=  \sum_{l=r}^{n}a_{rl}  /n, \text{ \  } \overline
 {a}_{..}=  \sum_{r,l=1}^{n}a_{rl}  /n^{2}$ and analogously for the terms
 $\overline{b}_{r.}$, $\overline{b}_{.l}$, and $\overline{b}$. where $\widehat{MDD}_{X}^{2}\left(  j\right) \geq 0 $ almost surely, for  $j\in\mathbb{N}^{+}$, where the equality holds under the condition that $E\left(  X_{t}%
|X_{t-j}\right)  =E\left(  X_{t}\right)  $.\footnote{It is important to remark that $\widehat{V}_{X}^{2}\left(j\right)  $ and $\widehat{MDD}_{X}^{2}\left(  j\right)  $ are not unbiased estimators of their respective population
counterparts.} In a similar manner, sample AMDC function is defined as%
\begin{equation}
	\widehat{MDC}_{X}^{2}\left(  j\right)  =\left\{
	\begin{array}
		[c]{c}%
		\frac{\widehat{MDD}_{X}^{2}\left(  j\right)  }{\sqrt{\widehat{Var}\left(  X_{t}\right)
				\widehat{V}^{2}\left(  0\right)  }},\text{ \ }\widehat{Var}\left(  X%
		_{t}\right)  \widehat{V}_{X}^{2}\left(  0\right)  \neq0;\\
		0,\text{ \ \ \ \ \ \ \ \ \ \ otherwise}
	\end{array}
	\right.
\end{equation}

 \noindent In order to develop theoretical results for the asymptotic properties of the statistic	in a time series framework, we need the following assumptions:
	
	\begin{enumerate}

		\item \ $\left\{  X_{t}\right\}  $ is a strictly stationary
		$\alpha$-mixing process with mixing coefficients $\alpha\left(  j\right)  ,$
		$j\geqslant1.$
		
		\item $E\left\vert X_{t}\right\vert ^{2}<\infty.$
		
		\item  The mixing coefficients of $\left\{  X_{t}\right\}  ,$
		$\alpha\left(  j\right)  ,$ satisfy
		
		\begin{itemize}
			\item $\sum_{j=-\infty}^{\infty}\alpha\left(  j\right)  <\infty;$
			
			\item $\alpha\left(  j\right)  =O\left(  1/j^{2}\right)  .$
		\end{itemize}
		
		\item  The kernel function $k\left(  .\right)  $ is defined
		such that $k:\mathbb{R\longrightarrow}\left[  -1,1\right]  $, is symmetric and
		continuous at zero and at all but a finite number of points. Moreover, it holds
		that $k\left(  0\right)  =1$, $\int_{-\infty}^{\infty}k^{2}\left(  z\right)
		dz<\infty$ and $\left\vert k\left(  z\right)  \right\vert \leq C\left\vert
		z\right\vert ^{-b}$ for large $z$ and $b>1/2$.
	\end{enumerate}
	
	The above assumptions are in part borrowed from \cite{fokianos2017consistent}
	even though we impose slightly more stringent conditions upon the moments of
	the random variables by requiring that $X_{t}$ has finite second moment. The first assumption about the series' mixing coefficients is necessary to
	show the consistency of the AMDD estimator by reason of the strong law of large
	numbers (SLLN) for $\alpha$-mixing random variables. The proof represents an
	elaboration of the original consistency theorem presented in \cite{szekely2007measuring} for the general case of distance covariance. 
	\begin{proposition}
		Suppose Assumptions (1) and (2) hold true. Then for all \ $j=0,\pm1,\pm2,...$we
		have%
		\[
		\widehat{MDD}_{X}^{2}\left(  j\right)  \overset{A.S.}{\longrightarrow}%
		MDD_{X}^{2}\left(  j\right)
		\]
		almost surely, as $n\rightarrow\infty$ $.$
	\end{proposition}
	
It is possible to reformulate the original test statistic from Equation (\ref{eq:hongtest}) as%
	\begin{equation}
		M_{n}=\sum_{j=1}^{n-1}\left(  n-j\right)  k^{2}\left(  j/p\right)
		\widehat{MDD}_{X}^{2}\left(  j\right)  .
	\end{equation}

 	 We can now formulate the following two theorems concerning the
	asymptotic behavior of our statistic. Both results are straightforwardly
	derived from \cite{hong1999hypothesis}. In particular, Hong is able to prove to elaborate a general proof of the theorems, meaning that the results are valid for any	possible form of the generalized spectrum and its derivatives: we can easily
	adapt his results to our framework. In particular we are able to define a standardized version of the
	test statistic that is asymptotically distributed as a standard normal random variable under the null hypothesis of independence.
	
	\begin{theorem}
		Suppose Assumptions 2 and 4 hold, and $p=cn^{\lambda}$, with $c>0$,
		$\lambda\in\left(  0,1\right)  .$ Then, assuming $X_{t}$ is independently and
		identically distributed%
		\[
		M_{n}=\left[  \sum_{j=1}^{n-1}\left(  n-j\right)  k^{2}\left(  j/p\right)
		\widehat{MDD}_{X}^{2}\left(  j\right)  -\widehat{C}_{0}\right]  \cdot\left(
		\widehat{D}_{0}\right)  ^{-1/2}%
		{d}{\longrightarrow}N\left(  0,1\right)
		\]
		as $n\longrightarrow\infty$, in distribution.
	\end{theorem}

	The terms used for the standardization, $\widehat{C}_{0}$ and
	$\widehat{D}_{0}$ are defined as follows%
	\begin{align}
		\widehat{C}_{0}  &  =\widehat{R}_{1}\left(  0\right)
		{\displaystyle\sum_{j=1}^{n-1}}
		k^{2}\left(  j/p\right)  \int\widehat{\sigma}_{0}\left(  v,-v\right)
		dW\left(  v\right) \\
		\widehat{D}_{0}  &  =2\widehat{R}_{1}^{2}\left(  0\right)
		{\displaystyle\sum_{j=1}^{n-2}}
		k^{4}\left(  j/p\right)  \int\left\vert \widehat{\sigma}_{0}\left(
		v,v^{\prime}\right)  \right\vert ^{2}dW\left(  v\right)  dW\left(  v^{\prime
		}\right)  ,\nonumber
	\end{align} and $\widehat{R}_{1}\left(  0\right)  =n^{-1}%
	{\displaystyle\sum_{t=1}^{n}}
	\left(  X_{t}-\overline{X}\right)  ^{2}$. We can also write:%
	
	\begin{equation}
		\int\sigma_{0}\left(  v,-v\right)  d\mathcal{W}\left(  v\right)  \approx
		\int\frac{1-\left\vert \phi\left(  v\right)  \right\vert ^{2}}{\pi\left\vert
			v\right\vert ^{2}}dv=E\left\vert X_{t}-X_{t}^{\prime}\right\vert
	\end{equation}

	and%
	\begin{align}
		&  \int\int\left\vert \widehat{\sigma}_{0}\left(  v,v^{\prime}\right)
		\right\vert ^{2}d\mathcal{W}\left(  v\right)  d\mathcal{W}\left(  v^{\prime
		}\right)   =\int_{\mathbb{R}^{2}}\frac{\left\vert \phi_{0}\left(  v,v^{\prime}\right)
			-\phi\left(  v\right)  \phi\left(  v^{\prime}\right)  \right\vert ^{2}}%
		{\pi^{2}\left\vert v\right\vert \left\vert v^{\prime}\right\vert }%
		dvdv^{\prime}  =V_{x}^{2}\left(  0\right)  ,
	\end{align} Finally, we report the following result:
	
	\begin{theorem}
		Suppose that Assumptions 1, 3(i), and 4 hold and $p=cn^{\lambda}$ for $c>0$
		and $\lambda\in\left(  0,1\right)  .$ Then ,%
		\[
		\frac{\sqrt{p}}{n}M_{n}\overset{P}{\longrightarrow}\frac{\frac{\pi}{2}%
			\int\int\left\vert f^{\left(  1,0\right)  }\left(  \omega,0,v\right)
			-f_{0}^{\left(  1,0\right)  }\left(  \omega,0,v\right)  \right\vert d\omega
			d\mathcal{W}\left(  v\right)  }{\left[  D_{0}\int_{0}^{\infty}k^{4}\left(
			z\right)  dz\right]  ^{1/2}}%
		\]
		as $n\longrightarrow\infty$, in probability.
	\end{theorem}

	\section{Simulations}
	
	\label{section:empirics}

	\noindent In this section, we carry out simulation experiments to assess
	the performance of the $M_{n}$ test in finite samples. The performance of the test is directly compared with that of the two test statistics introduced by \cite{escanciano2006generalized} and \cite{wang2022testing}, which we introduce in this section. 
	
	In the case of Wang et al.'s paper, the test statistic takes the formulation

	\begin{equation}
		M_{wn}^{F}\left(  p\right)  =n\sum_{j=1}^{p}\omega_{j}\left\vert \left\vert
		\widehat{MDDM}\left(  X_{t}|X_{t-j}\right)  \right\vert \right\vert _{F},
	\end{equation}
	and is based on the concept of the martingale difference divergence matrix, written
	here as $MDDM\left(  \cdot\right)  $ (\cite{lee2018martingale}). For the moment, suffice it to say that $MDDM$ represents a
	multivariate extension of the concept of martingale difference divergence and,
	in the univariate case, the identity $\widehat{MDDM}\left(  X_{t}%
	|X_{t-j}\right)  =\widehat{MDD}_{X}^{2}\left(  j\right)  $ holds. Here
	$\omega_{j}$ denotes the weight $\omega_{j}=\left(  n-j+1\right)  /\left(  n\ast
	j^{2}\right)  $ $\ $and $\left\vert \left\vert \cdot\right\vert \right\vert
	_{F}$ indicates the Frobenius norm. The test specification that we shall consider in our comparisons will therefore be%
	\begin{equation}
		M_{wn}^{F}\left(  p\right)  =n\sum_{j=1}^{p}\omega_{j}\left\vert \left\vert
		\widehat{MDD}_{X}\left(  j\right)  \right\vert \right\vert .
	\end{equation}

	Among all the different test specifications proposed by \cite{wang2022testing}, we
	choose $M_{wn}^{F}$ for conducting our comparisons as it appears to deliver the best
	compromise in terms of size and power properties. Even though the two competing test specifications $M_{n}$ and $M_{wn}^{F}$ are very similar in their formulation (they both correspond to a summation over different lags of the auto martingale difference divergence function), they originate from a different theoretical background, as Wang et al.'s statistic is not explicitly based on Hong's generalized spectral density approach. The original paper is mainly concerned with testing the MDH for multivariate regression residuals. In this sense, we consider only a simplified version of the methodology in our comparisons.

	Concerning Escanciano and Velasco's work, the construction of their statistic also has its foundations in Hong's approach; simultaneously, the paper departs from the original definition by using the concept of generalized spectral distribution function (borrowed from that of generalized spectral density). The test statistic takes the following form
	\begin{equation}
	D_{n}^{2}=\sum_{j=1}^{n-1}\left( n-j\right) \frac{1}{\left( j\pi \right) ^{2}%
	}\int_{\mathbb{R}}\left\vert \widehat{\sigma }_{j}\left( v\right)
	\right\vert ^{2}dW\left( v\right) 
	\end{equation}
	
	where $W\left(  \cdot\right)  $ is an absolutely continuous cumulative
	distribution function (CDF) with respect to the Lebesgue measure. We can notice that the above formulation is very similar to that of Equation (\ref{eq:hongtest}): the main difference, in this case, is the lack of use of a kernel function and a smoothing parameter. Moreover, the weighting function adopted in the definition of $D_{n}^{2}$ is assumed to be a proper probability distribution function, a detail that differentiates this statistic from the other two, both based on the assumption of a non-integrable weighting function.
	In particular, if $W\left(  \cdot\right)  $ is assumed to be the CDF of a standard normal random variable, $D_{n}^{2}$ takes the form
	\begin{equation}
	D_{n}^{2}=\sum_{j=1}^{n-1}\left( n-j\right) \frac{1}{\left( j\pi \right) ^{2}%
	}\sum_{t=j+1}^{n}\sum_{s=j+1}^{n}\left( X_{t}-\overline{X}_{n-j}\right)
	\left( X_{s}-\overline{X}_{n-j}\right) e^{-\frac{1}{2}\left(
		X_{t-j}-X_{s-j}\right) ^{2}}.
	\end{equation}
	Similarly to our statistic (in the case of which we have the use of a kernel function), $D_{n}^{2}$ considers all lags of the variable $X_t$ in its computation, with weight terms discounting high-order, lags severely; Wang et al.'s statistic, on the other hand, is characterized by a fixed truncation lag $p$ in its formulation.
	
	 Regarding the assumptions necessary to derive the asymptotic properties of the different statistics, no mixing or asymptotic independence assumption is needed in the case of \cite{escanciano2006generalized} and \cite{wang2022testing}. This differentiates their statistics from our case, which, being based on the work of \cite{hong1999hypothesis}, requires the definition of a proper mixing condition.

	 As the asymptotic null distribution of both $D_{n}^{2}$ and $M_{wn}^{F}$ is non-pivotal, the critical values of the two test statistics need to be approximated by bootstrap methodology. To this purpose, a wild bootstrap procedure (\cite{wu1986jackknife}; \cite{liu1988bootstrap}) is adopted. Specifically, the bootstrap sample $\left\{ X_{t}^{\ast }\right\} _{t=1}^{n}$ is generated according to the formula $X_{t}^{\ast }=X_{t}w_{t}^{\ast }$ with $\left\{ w_{t}^{\ast }\right\}$ being a sequence of iid
	 random variables with mean zero, unit variance, and bounded support. For instance, \cite{wang2022testing} choose $\left\{ w_{t}^{\ast }\right\} $ as iid random variables with the
	 distribution (\cite{mammen1993bootstrap})%
	 \begin{eqnarray}
	 	P\left( w_{t}^{\ast }=\frac{1-\sqrt{5}}{2}\right)  &=&\frac{\sqrt{5}+1}{2%
	 		\sqrt{5}}, \\
	 	P\left( w_{t}^{\ast }=\frac{1+\sqrt{5}}{2}\right)  &=&\frac{\sqrt{5}-1}{2%
	 		\sqrt{5}}  \nonumber
	 \end{eqnarray}
	 Alternatively, Escanciano and Velasco adopt a similar approach to bootstrap by choosing iid $\left\{ w_{t}^{\ast
	 }\right\} $ having Rademacher distribution  
	\begin{equation}
		P\left( w_{t}^{\ast }=1\right) =P\left( w_{t}^{\ast }=-1\right) =0.5 		
	\end{equation}
as discussed in the paper of \cite{liu1988bootstrap}.

	\subsection*{Bootstrap Procedure}
	In the present section, we introduce the wild bootstrap methodology that is adopted throughout the simulated experiment to approximate the asymptotic distribution of the test statistic $M_{n}$. Our approach is based on the methodology proposed in \cite{fokianos2018testing}, which offers a simple and efficient way to approximate the limit distribution of degenerate U-and V-statistics, and relies on the generation of a sequence of auxiliary random
	variables $\left\{  W_{t,n}^{\ast}\right\}  _{t=1}^{n-\left\vert j\right\vert
	}$ for obtaining the bootstrap statistic. The procedure was originally described in \cite{leucht2013dependent} for
    the case-dependent data and was based on the assumption that the auxiliary random
	variables $W_{t,n}^{\ast}$ followed a first-order autoregressive model. This peculiar feature made it suitable to consider a null hypothesis that allowed some form of dependence in the data. 
	
	Fokianos and Pitsillou, on the contrary, by focusing on the hypothesis of serial
	independence, found it convenient to simply assume $\left\{  W_{t,n}^{\ast}\right\}
	_{t=1}^{n-\left\vert j\right\vert }$ to be a sequence of i.i.d. standard normal
	random variables.\newline In this paper, we shall explore both the latter
	approach as well as the original methodology as described in Leucht and
	Neumann. In particular, the original assumption concerning the existence of
	some form of dependence in the auxiliary variables should be more
	appropriate in our framework, considering that our focus is to test for the null hypothesis of conditional mean independence. 
	
	Analogously to Leucht and Neumann, we need the following set of assumptions
	
	\begin{itemize}
		\item $\left\{  X_{t}\right\}  $ is a strictly stationary time, $\tau
		$-dependent process 
		
		\item The variables $\left\{  W_{t,n}^{\ast}\right\}  _{t=1}^{n}$ follow an
		$AR\left(  1\right)  $ process:
\begin{equation}
		W_{t,n}^{\ast}=e^{-1/l_{n}}W_{t-1,n}^{\ast}+\sqrt{1-e^{-2/l_{n}}}
		\varepsilon_{t}^{\ast},
\end{equation}

		where $W_{0,n}^{\ast},\varepsilon_{1}^{\ast},...,\varepsilon_{n}^{\ast}$ are
		independent standard normal variables. The parameter $l_{n}$ is required to
		satisfy the condition $l_{n}=o\left(  n\right)  $ $\ $and $l_{n}
		\rightarrow_{n\rightarrow\infty}\infty.$ This leaves some freedom in its
		choice. In particular, when $l_{n}=0,$ we are in the case analysed
		by Fokianos, as $W_{t,n}^{\ast}\rightarrow\varepsilon_{t}^{\ast}$ and the sequence of auxiliary variables is entirely made of independent standard
		normal variables.
	\end{itemize}

	Remember now that sample martingale difference divergence takes the form
	$\ \widehat{MDD}_{X}^{2}\left(  j\right)  =\left(  n-\left\vert j\right\vert
	\right)  ^{-2}\sum_{r,l=1+\left\vert j\right\vert }^{n}A_{rl}B_{rl}$ where,
	based on the sample $\left\{  X_{t},X_{t-\left\vert j\right\vert
	}:t=1+\left\vert j\right\vert ,...,n\right\}  , $ we calculate the double
	centered Euclidean distance matrices $A=\left(  A_{rl}\right)  $ and
	$B=\left(  B_{rl}\right)  $. If we define the $\left(  \left(  n-\left\vert
	j\right\vert \right)  \times1\right)  $ unitary vector as $e_{n-\left\vert
		j\right\vert }$ we can equivalently express the statistic as%
	\begin{equation}
		\widehat{MDD}_{X}^{2}\left(  j\right)  =\left(  n-\left\vert j\right\vert
		\right)  ^{-2}\left(  e_{n-\left\vert j\right\vert }\right)  ^{T}\ast A\circ
		B\ast\left(  e_{n-\left\vert j\right\vert }\right)
	\end{equation}
	where $\circ$ stands for the element-wise product operator for two matrices.
	\newline We can estimate the distribution of the test statistic $M_{n}$ by
	that of
	\begin{equation}
		M_{n}^{\ast}=\sum_{j=1}^{n-1}\left(  n-j\right)  k^{2}\left(  j/p\right)
		\frac{1}{\left(  n-\left\vert j\right\vert \right)  ^{2}}\left(
		W_{n-\left\vert j\right\vert }^{\ast}\right)  ^{T}\ast A\circ B\ast\left(
		W_{n-\left\vert j\right\vert }^{\ast}\right)
	\end{equation}
	where $W_{n-\left\vert j\right\vert }^{\ast}$ is the $\left(  \left(
	n-\left\vert j\right\vert \right)  \times1\right)  $ auxiliary variables'
	vector. For a number of bootstrap replicates equal to $B$, the bootstrap
	$p$-values are then obtained as the ratio $\sum_{b=1}^{B}\mathbb{I}\left(
	T_{n}^{\ast}\geqslant T_{n}\right)  /B,$ where $\mathbb{I}\left(
	\cdot\right)  $ is the indicator function. In our simulations, we shall use
	both independent ($l_{n}=0$) and dependent $\left(  l_{n}\neq0\right)  $ wild
	bootstrap procedures to compute empirical $p$-values.
	
	\subsection*{Results}

	In our experiment, we consider a number of different DGPs to study the behavior of the test statistic under the null and the alternative hypothesis. These DGPs were originally introduced in \cite{escanciano2006generalized}'s paper; in the sequel $\varepsilon_{t}$ and $u_{t}$ will be independent sequences of i.i.d. N(0,1). The models used in the simulations include respectively five MDS:
	
	\begin{itemize}
		\item \textbf{DGP (1) } Independent and identically distributed N(0,1) variates (IID).
		\item \textbf{DGP (2,3,4) } GARCH(1,1) processes, $X_{t}=\varepsilon_{t}\sigma_{t}$, with $\sigma_{t}^{2}=0.001+\alpha X_{t-1}^{2}+ \beta \sigma_{t-1}^{2}$, and the following combinations for $\left(\alpha, \beta \right):\left(0.01, 0.97\right),\left( 0.09, 0.89\right) \text{and} \left(0.09, 0.90\right)$.
		\item  \textbf{DGP (5) } Stochastic volatility model, $X_{t}=\varepsilon\exp\left(  \sigma_{t}\right) $, with $\sigma_{t}= \alpha \sigma_{t-1}+ \beta u_{t}$, and $\left(\alpha, \beta \right):\left(0.936, 0.32\right)$. 
	\end{itemize}
		And the following non-martingale difference sequences:
	\begin{itemize}
		\item  \textbf{DGP (6) } Threshold autoregressive model of order one :
		\begin{equation}
				X_{t} = -1.5X_{t-1}+\varepsilon_{t}\text{ if }X_{t-1}<0 \text{ and }  X_{t} = 0.5X_{t-1}+\varepsilon_{t},\text{if }X_{t-1}\geq0 \nonumber
		\end{equation}
		\item  \textbf{DGP (7,8) } Bilinear Processes: 	$X_{t}=\varepsilon_{t}+ \alpha \varepsilon_{t-1}X_{t-1}+ \beta \varepsilon_{t-1}X_{t-2},$
		with parameter values $\left(\alpha, \beta \right):\left(0.15, 0.05\right) \text{and} \left(0.25, 0.25\right)$.
		\item  \textbf{DGP (9) } Non-linear moving average model : $X_{t}= -0.6 \varepsilon^{2}_{t-1} + \varepsilon_{t} $.
		\item  \textbf{DGP (10) } Non-linear moving average model : $X_{t}= \varepsilon_{t-1} \varepsilon_{t-2} \left( \varepsilon_{t-2}  + \varepsilon_{t} + 1 \right) $.
	\end{itemize}

For all DGPs, we consider sample sizes of $n = 100$ and 300 and set the significance level to $5\%$. The number of Monte Carlo experiments will be  1000. The number of bootstrap replications is set to be $B = 499$ for all test statistics, where for each one, we follow the bootstrap methodologies presented in previous sections. 

Concerning the  formulation for test statistic $M_{n}$, results are obtained by using both Parzen kernel (PAR), and Bartlett kernel (BAR). To examine the sensitivity of the test statistic $M_{n}$ on the values of bandwidth $p$, we shall use $p=n^{\lambda}$ with $\lambda = 1/5, \text{ \\ } 2/5,$ and $ \text{ \\ } 3/5$ throughout the experiment. 
In addition to this, we will separately consider the case in which $l_{n}=0$ (wild independent bootstrap) and $l_{n}=7$ (wild dependent bootstrap)\footnote{We report the results obtained for different $l_{n}$s in separate tables, where it is explicitly mentioned when $l_{n}=7$}. In the case of the statistic $M^{F}_{wn}$ the maximum lag parameter will be assumed to take values $p=1,3$ and 6. 

From Table \ref{tab:TabSiz1}, we can observe that the different specifications of $M_{n}$ offer a satisfactory size performance in most cases, even though they tend to be slightly undersized for DGP (5). For this specific DGP, other test specifications appear to provide a better size approximation. Good results, however, are still obtained by setting $\lambda=3/5$, for both Parzen and Bartlett kernels.

Table \ref{tab:TabSize2} reports results obtained for different specifications of $M_{n}$ when $l_{n}=7$. Results clearly show how assuming some form of dependence in the auxiliary bootstrap variables has a negative impact on the test's size performance. In particular, $M_{n}$ appears to be persistently undersized.

Table \ref{tab:TabPow1} shows the empirical power of the statistics against DGPs (6) to (10). We can notice that, in most of the cases examined, the power of $M_{n}$ decreases as the value of $\lambda$ increases. 
 Results are largely similar when using the Bartlett and Parzen kernel: the main difference in this sense can be spotted when considering DGP (9), for which results obtained for the Bartlett kernel appear to be inferior for a small sample size. Empirical power levels rapidly approach one for DGPs (6), (8), and (9) as a larger sample is considered. This is true for all specifications of $M_{n}$ except for the case $\lambda=3/5$. For these DGPs, the performance observed in the different statistics is comparable, with no test among $M_{n}$, $M^{F}_{wn}$ and $D^2_{n}$ neatly outperforming the others. The statistic $M_{n}$ generally exhibits low empirical power against DGPs (7) and (10): this is true for both kernel specifications and for any value of $\lambda$, even though choosing larger values of $\lambda$ seems to have a positive impact. For these two DGPs, the two competing statistics, $M^{F}_{wn}$ and $D^2_{n}$, seem to display better results. Empirical power levels of the statistic $M^{F}_{wn}$ seem to be only marginally affected by the maximum lag parameter $p$.

Finally, Table \ref{tab:TabPow2} reports the empirical power levels computed for the $M_{n}$ when $l_{n}=7$.
In this case, all specifications of the test have low empirical power. In general, we can conclude that a wild independent bootstrap methodology ($l_{n}=0$) should be preferred.

	\section{Empirical Applications}
	
\begin{figure}[ptb]
		\centering
		\begin{subfigure}[b]{0.49\textwidth}
			\centering
			\includegraphics[width=\textwidth]{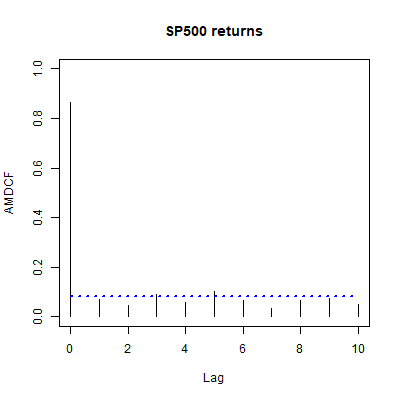}
		\end{subfigure}
		\begin{subfigure}[b]{0.49\textwidth}
			\centering
			\includegraphics[width=\textwidth]{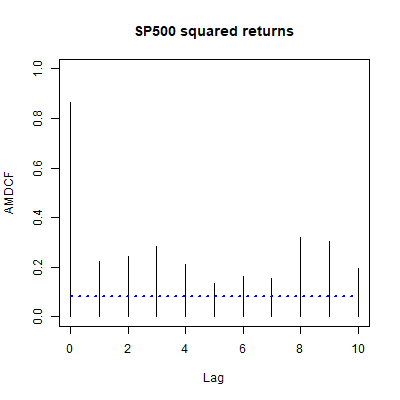}
		\end{subfigure}
		\par
		\begin{subfigure}[b]{0.49\textwidth}
			\centering
			\includegraphics[width=\textwidth]{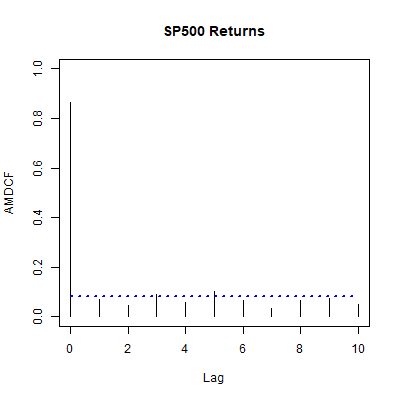}
		\end{subfigure}
		\begin{subfigure}[b]{0.49\textwidth}
			\centering
			\includegraphics[width=\textwidth]{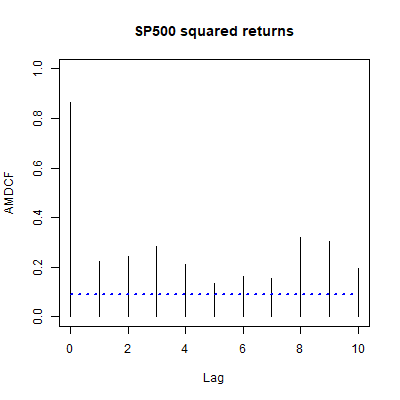}
		\end{subfigure}
		\caption{The AMDCF plot of S\&P returns and squared returns.\newline Upper
			panel: Critical values computed through wild independent bootstrap. \newline
			Lower panel: Critical values computed through wild dependent bootstrap.}%
		\label{fig:amdcf ret sqret}%
	\end{figure}

	\smallskip Along the lines of \cite{fokianos2017consistent}, we study the monthly excess
	returns series of the S\&P 500 index starting from January 1926 and extending
	up until December 1991 (\cite{tsay2005analysis}). In the paper (following the original
	inquiry of \cite{zhou2012measuring}), it is shown that while the original series exhibits
	only moderate signs of autocorrelation, the series of the squared returns are
	characterized by strong linear dependence (Figure \ref{spSEries}). This is one
	feature that is common to many financial return time series and suggests the
	existence of some nonlinear dependence structures. The auto distance
	covariance function (ADCF) plots are then used to
	investigate the existence of nonlinear serial dependence. In the example, the
	critical values are computed using a wild (independent) bootstrap procedure
	(in the original papers, critical values were computed by adopting a subsampling
	procedure). We also report the auto martingale difference divergence (AMDCF) plots in Figure \ref{fig:amdcf ret sqret}, where the upper
	plots' critical values are computed using a wild-independent bootstrap
	procedure and those in the lower plots using the wild-dependent bootstrap
	defined in our paper.\newline As we can notice, both ADCF and AMDCF
	plots signal the existence of some form of dependence, even if the evidence is
	much weaker in the case of the AMDCF plot. In Table \ref{tab:rest} we have
	reported the results obtained for the testing methodologies $T_{n}$ and
	$M_{n}$ for values of the smoothing parameter $p=4,15,$ and $55$. As we can
	see, the hypothesis that the series are a MDS is not rejected in all
	instances.: nonlinear dependence is not found in the conditional mean and has
	to be searched for in other moments.
	\section*{Conclusions}
	
	We have introduced a novel test for the martingale difference
	hypothesis based on the martingale difference divergence function.
	In the context of generalized spectral density theory\ we have seen how
	martingale difference divergence allows us to find an explicit solution to the
	weighted quadratic norm of a form of generalized covariance: this is possible
	by expressing the weight function in the integral in the original test of \cite{hong1999hypothesis} as the non-integrable weight function introduced in \cite{szekely2007measuring}. In this sense, we are able to define a dependence test largely
	analogous to that introduced by \cite{fokianos2017consistent} to test the
	hypothesis of general serial independence. We have therefore shown how our
	martingale test can be seen, together with the independence test of Fokianos
	and Pitsillou, as part of a family of statistics that can be obtained from
	generalized spectral density theory.\newline In particular, we have
	demonstrated how the two dependence tests can be used jointly in order to
	better characterize the nature of the serial dependence found in the series
	(linear or nonlinear, conditional mean or in other higher moments). Hong
	proved how it is possible to obtain a family of tests addressed to the
	analysis of different aspects of temporal dependence starting from the
	derivatives of various orders of the generalized spectrum. In this sense, one
	might think that suitable variations made to the distance covariance statistic
	could allow the construction of new forms of tests based on different degrees
	of derivatives of the generalized spectrum.
	
	Another possible extension of our testing methodology could be to adapt it to test
	for the martingale difference hypothesis in multivariate time series. In this
	sense, we could follow \cite{fokianos2018testing} in the construction of a
	test for multivariate serial dependence based on generalized spectral
	density theory.
	\newline Concerning its large sample
	performance, our statistic shows the good empirical size and power
	properties, being capable of competing against the test\ for martingale
	difference hypothesis recently introduced by \cite{wang2022testing} when an appropriate choice for the kernel and the smoothing parameter is made. Indeed, the two statistics share remarkable similarities, as they both make use of the concept of martingale difference divergence function in their definition.  Despite their common features, however, we argue that the introduction of our test statistic can still provide a relevant contribution due to its peculiar theoretical construction and the fact that it can be regarded as part of a larger family of statistics looking at different aspects of serial dependence.

\bibliography{references}
\clearpage
	
	\section{Tables}

	\vspace{1cm}
		
	\begin{table}[ptbh]
		\caption{Size ($ \times  100$) of all tests for DGPs from (1) to (5) at level 5$\%$}%
		\label{tab:TabSiz1}%
		\renewcommand*{\arraystretch}{1.2} \centering
		\vspace{0.05cm} \scalebox{0.825}{\begin{tabular}{cccccccccccccccccccccc}
				\toprule
				\toprule
							
				&       & \multicolumn{3}{c}{\textbf{DGP1}} &  &     \multicolumn{3}{c}{\textbf{DGP2}} & &\multicolumn{3}{c}{\textbf{DGP3}}& & \multicolumn{3}{c}{\textbf{DGP4}}&& \multicolumn{3}{c}{\textbf{DGP5}}&\\
				\cmidrule{3-5}     \cmidrule{7-9}   \cmidrule{11-13}    \cmidrule{15-17} \cmidrule{19-21}
				
				n &   & 100  &   & 300  &  &100 & &300 & &100&&300&&100&&300& &100&&300&\\
				\hline
				
				$M_{n},(Par),\lambda=1/5$ &  & 5.1 &  & 5.0 &  & 4.1& & 4.6 &  &4.5&&4.9&  &3.8&&5.5&  &2.6&&3.5&\\
				$M_{n},(Par),\lambda=2/5$ &  & 4.5 &  & 5.0 &  &5.1& &5.9    &  &4.1&&6.1&  &4.2&&4.4&  &3.8&&4.6&\\
				$M_{n},(Par),\lambda=3/5$ &  & 5.0 &  & 5.9 &  &6.4& &4.9    &  &5.2&&5.9&  &5.6&&6.5&  &4.1&&5.4& \\
				\hline
				$M_{n},(Bar),\lambda=1/5$ &  & 5.0 &  & 3.9 &  &4.8& &5.1&  &4.9&&4.9&  &4.9&&4.9&  &4.2&&3.1& \\
				$M_{n},(Bar),\lambda=2/5$ &  & 4.9 &  & 6.2 &  &4.6& &4.7&  &6.4&&5.6&  &4.4&&5.3&  &2.7&&3.7& \\
				$M_{n},(Bar),\lambda=3/5$ &  & 4.5 &  & 6.1 &  &4.9& &6.1&   &6.2&&4.6&  &5.1&&5.5&  &4.0&&4.8&\\
				\hline
				$M^{F}_{wn}, p=1$        &  & 5.0 &  & 4.7 &  &5.7&  &3.8&   &7.0&&5.3&  &3.9&&5.1&  &5.2&&5.2&\\
				$M^{F}_{wn}, p=3$         &  & 4.4 &  & 4.7 &  &4.8&  &4.6&   &4.2&&6.1&  &4.6&&5.1&  &5.6&&4.9&\\
				$M^{F}_{wn}, p=6$         &  & 5.3 &  & 4.7 &  &6.6&  &5.6&   &6.0&&5.6&  &6.3&&5.8&  &5.3&&4.1&\\
				\cline{1-21} \vspace{1.6mm}
				
				$D^{2}_{n}$    &  & 5.4 &  & 4.1   & &6.8&  &4.3&   &5.9&&4.3& &4.8&&5.8&  &5.3&&5.2&\\

				\bottomrule
		\end{tabular}}
	
	\end{table}

	\bigskip
	\vspace{3cm}

	\begin{table}[ptbh]
		\caption{Size ($ \times 100$) of $M_{n}$, $l_{n}=7$, for DGPs from (1) to (5) at level 5$\%$}%
		\label{tab:TabSize2}%
		\renewcommand*{\arraystretch}{1.2} \centering
		\vspace{0.05cm} \scalebox{0.825}{\begin{tabular}{cccccccccccccccccccccc}
				\toprule
				\toprule

				&       & \multicolumn{3}{c}{\textbf{DGP1}} &  &     \multicolumn{3}{c}{\textbf{DGP2}} & &\multicolumn{3}{c}{\textbf{DGP3}}& & \multicolumn{3}{c}{\textbf{DGP4}}&& \multicolumn{3}{c}{\textbf{DGP5}}&\\
				\cmidrule{3-5}     \cmidrule{7-9}   \cmidrule{11-13}    \cmidrule{15-17} \cmidrule{19-21}
				
				n &   & 100  &   & 300  &  &100 & &300 & &100&&300&&100&&300& &100&&300&\\
				\hline

				$M_{n},(Par),\lambda=1/5$  &  & 1.2 &  & 3.3 &  &0.9& &2.9&  &1.0&&2.1&  &0.5&&1.5&  &1.0&&2.6&\\
				$M_{n},(Par),\lambda=2/5$ &  & 1.7 &  & 4.2 &  &0.2& &3.6&  &1.0&&1.5&  &0.5&&1.6&  &0.5&&2.2&\\
				$M_{n},(Par),\lambda=3/5$ &  & 1.9 &  & 4.2 &  &1.6& &4.4&   &1.5&&2.9&  &1.0&&2.9&  &1.1&&2.6&\\
				\hline
				$M_{n},(Bar),\lambda=1/5$ &  & 1.1 &  & 3.9 &  &0.7& &3.8&   &0.9&&2.2&  &0.4&&1.9&  &0.4&&1.8&\\
				$M_{n},(Bar),\lambda=2/5$ &  & 1.2 &  & 3.7 &  &1.4& &3.2&   &1.2&&2.0&  &0.1&&1.2&  &0.3&&2.3&\\
				$M_{n},(Bar),\lambda=3/5$&  & 1.7 &  & 3.6 &  &1.4& &3.7&   &1.6&&3.8&  &0.2&&2.3&  &0.9&&2.7&\\
				\bottomrule
		\end{tabular}}
		
	\end{table}

	\newpage
	
	\begin{table}[ptbh]
		\caption{The empirical power ($ \times 100$) of all tests for DGPs from (6) to (10) at level 5$\%$}%
		\label{tab:TabPow1}%
		\renewcommand*{\arraystretch}{1.1} \centering
		\vspace{0.05cm} \scalebox{0.825}{\begin{tabular}{cccccccccccccccccccccc}
				\toprule
				\toprule

				&       & \multicolumn{3}{c}{\textbf{DGP6}} &  &     \multicolumn{3}{c}{\textbf{DGP7}} & &\multicolumn{3}{c}{\textbf{DGP8}}& & \multicolumn{3}{c}{\textbf{DGP9}}&& \multicolumn{3}{c}{\textbf{DGP10}}&\\
				\cmidrule{3-5}     \cmidrule{7-9}   \cmidrule{11-13}    \cmidrule{15-17} \cmidrule{19-21}
				
				n &   & 100  &   & 300  &  &100 & &300 & &100&&300&&100&&300& &100&&300&\\
				\cmidrule{1-21}
				
				$M_{n},(Par),\lambda=1/5$ &  & 73 &   & 100 &  &11 & &44 &  &28 & &90&  &70& &100&    &15 & &28&\\
				$M_{n},(Par),\lambda=2/5$&  & 63 &   & 100 &  &11 & &25 &  &23 & &76&  &57& & 99&    &12 & &19&\\
				$M_{n},(Par),\lambda=3/5$ &  & 39 &   & 94  &  &9.3& &16 &  &16 & &45&  &40& & 88&    &8.0& &17&\\
				\cmidrule{1-21}
				$M_{n},(Bar),\lambda=1/5$ &  & 72 &   & 100 &  &12 & &44 &  &31 & &93&  &44& &100&    &6.2& &16&\\
				$M_{n},(Bar),\lambda=2/5$ &  & 67 &   & 100 &  &10 & &31 &  &23 & &83&  &37& & 98&    &6.1& &14&\\
				$M_{n},(Bar),\lambda=3/5$ &  & 48 &   & 98  &  &8.0& &21 &  &20 & &58&  &27& & 90&    &3.7&&8.1&\\
				\cmidrule{1-21}
				$M^{F}_{wn}, p=1$        &  & 48 &   & 100 &  &23 & &58 &  &22 & &89&  &44& &100&    &24 & &40&\\
				$M^{F}_{wn}, p=3$       &  & 48 &   & 100 &  &22 & &56 &  &23 & &89&  &38& &100&    &26 & &37&\\
				$M^{F}_{wn}, p=6$       &  & 100&   & 100 &  &17 & &58 &  &24 & &89&  &37& & 99&    &25 & &37&\\
				\cmidrule{1-21}\vspace{1.6mm}
				$D^{2}_{n}$    &  & 93 &   & 100 &  &20 & &64&   &54 & &98&  &89& &100&   &18 & &43&\\
				
				\bottomrule\bottomrule
		\end{tabular}}
	
	\end{table}
	
	\vspace{2cm}

	\begin{table}[ptbh]
		\caption{The empirical power ($ \times 100$) of $M_{n}$, $l_{n}=7$, for DGPs from (6) to (10) at level 5$\%$}%
		\label{tab:TabPow2}%
		\renewcommand*{\arraystretch}{1.1} \centering
		\vspace{0.05cm} \scalebox{0.825}{\begin{tabular}{cccccccccccccccccccccc}
				\toprule
				\toprule

				&       & \multicolumn{3}{c}{\textbf{DGP6}} &  &     \multicolumn{3}{c}{\textbf{DGP7}} & &\multicolumn{3}{c}{\textbf{DGP8}}& & \multicolumn{3}{c}{\textbf{DGP9}}&& \multicolumn{3}{c}{\textbf{DGP10}}&\\
				\cmidrule{3-5}     \cmidrule{7-9}   \cmidrule{11-13}    \cmidrule{15-17} \cmidrule{19-21}
				
				n &   & 100  &   & 300  &  &100 & &300 & &100&&300&&100&&300& &100&&300&\\
				\cmidrule{1-21}

				$M_{n},(Par),\lambda=1/5$ &  & 43 &   & 100 &  &3.2& &33 &  &8.8& &82&  &53& &100&    &6.0& &18&\\
				$M_{n},(Par),\lambda=2/5$ &  & 34 &   & 100 &  &2.6& &20 &  &6.9& &62&  &39& &9.0&    &3.0& &13&\\
				$M_{n},(Par),\lambda=3/5$ &  & 20 &   & 92  &  &3.2& &12 &  &5.4& &36&  &25& & 97&    &2.0&&8.0&\\
				\cmidrule{1-21}
				$M_{n},(Bar),\lambda=1/5$ &  & 44 &   & 100 &  &2.0& &32 &  &9.1& &80&  &17& &99 &    &0.8& &10&\\
				$M_{n},(Bar),\lambda=2/5$ &  & 38 &   & 100 &  &3.0& &26 &  &6.9& &70&  &16& & 98&    &0.5&&7.5&\\
				$M_{n},(Par),\lambda=3/5$ &  & 27 &   & 97  &  &3.0& &14 &  &6.2& &47&  &11& & 85&    &0.4&&4.7&\\

				\bottomrule\bottomrule
		\end{tabular}}
		
	\end{table}
	
\vspace{2cm}
	\begin{table}[ht]
		\centering
		\begin{tabular}
			[c]{c|c|c|c|c}\hline
			p & $T_{n}$ & $M_{n}$ & $M_{n} (l_{n}=7)$ & $M_{n} (l_{n}=15)$\\\hline
			4 & 0.014 & 0.86 & 0.842 & 0.804\\
			15 & 0.000 & 0.284 & 0.284 & 0.232\\
			55 & 0.008 & 0.286 & 0.266 & 0.196\\\hline
		\end{tabular}
		\caption{P-values for statistics $T_{n}$ and $M_{n}$ for S$\&$P 500 return
			series.}%
		\label{tab:rest}%
	\end{table}


	\newpage

\clearpage

\section*{Technical Appendix}
Let us first recall the main assumptions made in the paper:

\medskip

\emph{Assumption 1}. $\left\{  X_{t},\text{ }t\in\mathbb{Z}\right\}  $ a
univariate strictly stationary $\alpha$- mixing process with mixing
coefficients $\alpha\left(  j\right)  ,$ $j\geq1$ .

\emph{Assumption 2. }$E\left\vert X_{t}\right\vert ^{2}<\infty$

\emph{Assumption 3. }The mixing coefficients of $\left\{  X_{t}\right\}  $,
$\alpha\left(  j\right)  $, satisfy $\left(  i\right)  $ $\sum_{j=-\infty
}^{\infty}\alpha\left(  j\right)  <\infty$, $\left(  ii\right)  $
$\alpha\left(  j\right)  =O\left(  1/j^{2}\right)  $

\emph{Assmption 4. }\ Suppose that $k\left(  .\right)  $ is a kernel function
such that $k:\mathbb{R\rightarrow}\left[  -1,1\right]  $, is symmetric and is
continuous at 0 and at all but a finite number of points, with $k\left(
0\right)  =1$, $\int_{-\infty}^{\infty}k^{2}\left(  z\right)  dz<\infty$ and
$\left\vert k\left(  z\right)  \right\vert \leqslant C\left\vert z\right\vert
^{-b}$ for large $z$ and $b>1/2.$

\bigskip
In order to prove the first two lemmas of this paper, we need to define some
additional variables. First of all, we define the pseudoestimator%
\[
\overline{f}_{n}\left(  \omega,v\right)  =\frac{1}{2\pi}\sum_{j=-\left(
n-1\right)  }^{\left(  n-1\right)  }k\left(  j/p\right)  \left(  1-\left\vert
j\right\vert /n\right)  ^{1/2}\widetilde{\sigma}_{j}\left(  v\right)
e^{-ij\omega}%
,\]
where%
\[
\widetilde{\sigma}_{j}\left(  v\right)  =\frac{1}{n-\left\vert j\right\vert }%
{\displaystyle\sum_{t=\left\vert j\right\vert +1}^{n}}
\psi_{t}\psi_{t-\left\vert j\right\vert }\left(  v\right)
,\]
and%
\begin{align*}
\psi_{t}  &  =X_{t}-E\left(  X_{t}\right), \\
\psi_{t-\left\vert j\right\vert }\left(  v\right)   &  =e^{ivX_{t-\left\vert
j\right\vert }}-\phi\left(  v\right).
\end{align*}
We shall recall the following lemma contained in Szekely et al. (2007):
\begin{lemma}
[Szekely et al. (2007)]\textbf{\ }If $0<\alpha<2$, then for all $x$ in $%
\mathbb{R}
^{d}$
\[
\int_{
\mathbb{R}
^{d}}\frac{1-\cos\left\langle z,x\right\rangle }{\left\vert z\right\vert
_{d}^{d+\alpha}}dz=C\left(  d,\alpha\right)  \left\vert x\right\vert
^{\alpha}
,\]
where
\[
C\left(  d,\alpha\right)  =\frac{2\pi\Gamma\left(  1-\alpha/2\right)  }
{\alpha2^{\alpha}\Gamma\left(  \left(  d+\alpha\right)  /2\right)  }
,\]
and $\Gamma\left(  \cdot\right)  $ is the complete gamma function.
\end{lemma}

\bigskip

In particular, taking $\alpha=1,$ for the univariate case we have $C\left(
1,1\right)  =\pi.$ Accordingly%
\[
\int_{
\mathbb{R}
}\frac{1-\cos\left(  z\cdot x\right)  }{\left\vert z\right\vert ^{2}}%
dz=\pi\cdot\left\vert x\right\vert
.\]

\subsection*{Proofs}
\begin{lemma}
Suppose that $\left\{  X_{t},\text{ }t>1\right\}  $ satisfies \ Assumption 1
and 3(ii). Then,
\[
\left(  n-j\right)  ^{2}E\left\vert \widehat{\sigma}_{j}\left(  v\right)
-\widetilde{\sigma}_{j}\left(  v\right)  \right\vert ^{2}\leqslant C,
\]
and
\[
\left(  n-\left\vert j\right\vert \right)  E\left\vert \widetilde{\sigma}
_{j}\left(  v\right)  \right\vert ^{2}\leqslant C
\]
(where $C$ is a generic constant) uniformly in $v\in\mathbb{R}$. As hown in
Hong (1999) the result is also true under independence.
\end{lemma}

\begin{proof}
[Proof of Lemma 2]We can re-express the term $\widetilde{\sigma}_{j}\left(
v\right)  $ as%
\begin{align*}
\widetilde{\sigma}_{j}\left(  v\right)   &  =\frac{1}{n-\left\vert
j\right\vert }\sum_{t=\left\vert j\right\vert +1}^{n}\psi_{t}\psi
_{t-\left\vert j\right\vert }\left(  v\right) \\
&  =\frac{1}{n-\left\vert j\right\vert }\sum_{t=\left\vert j\right\vert
+1}^{n}\left(  X_{t}-E\left(  X_{t}\right)  \right)  \left(
e^{ivX_{t-\left\vert j\right\vert }}-\phi\left(  v\right)  \right) \\
&  =\frac{1}{n-\left\vert j\right\vert }\sum_{t=\left\vert j\right\vert
+1}^{n}X_{t}e^{ivX_{t-\left\vert j\right\vert }}-\frac{\phi\left(  v\right)
}{n-\left\vert j\right\vert }\sum_{t=\left\vert j\right\vert +1}^{n}
X_{t}-\frac{1}{n-\left\vert j\right\vert }E\left(  X_{t}\right)
\sum_{t=\left\vert j\right\vert +1}^{n}e^{ivX_{t-\left\vert j\right\vert }
}+E\left(  X_{t}\right)  \phi\left(  v\right)
\end{align*}

Subtracting $\widetilde{\sigma}_{j}\left(  v\right)  $ from $\widehat{\sigma
}_{j}\left(  v\right)  $ we get:%
\begin{align*}
&  \widehat{\sigma}_{j}\left(  v\right)  -\widetilde{\sigma}_{j}\left(
v\right)  =\\
&  =-\frac{1}{\left(  n-\left\vert j\right\vert \right)  ^{2}}\sum
_{t=\left\vert j\right\vert +1}^{n}\psi_{t}\sum_{t=\left\vert j\right\vert
+1}^{n}\psi_{t-\left\vert j\right\vert }\left(  v\right)  \\
&  =-\frac{1}{\left(  n-\left\vert j\right\vert \right)  ^{2}}\sum
_{t=\left\vert j\right\vert +1}^{n}\left(  X_{t}-E\left(  X_{t}\right)
\right)  \sum_{t=\left\vert j\right\vert +1}^{n}\left(  e^{ivX_{t-\left\vert
j\right\vert }}-\phi\left(  v\right)  \right)
\end{align*}

The Cauchy-Schwarz inequality implies that%
\begin{align*}
&  E\left\vert \widehat{\sigma}_{j}\left(  v\right)  -\widetilde{\sigma}
_{j}\left(  v\right)  \right\vert ^{2}=\\
&  =E\left\vert \frac{1}{\left(  n-\left\vert j\right\vert \right)  ^{2}}
\sum_{t=\left\vert j\right\vert +1}^{n}\psi_{t}\sum_{t=\left\vert j\right\vert
+1}^{n}\psi_{t-\left\vert j\right\vert }\left(  v\right)  \right\vert ^{2}\\
&  =\frac{1}{\left(  n-\left\vert j\right\vert \right)  ^{4}}E\left\vert
\sum_{t=\left\vert j\right\vert +1}^{n}\psi_{t}\sum_{t=\left\vert j\right\vert
+1}^{n}\psi_{t-\left\vert j\right\vert }\left(  v\right)  \right\vert ^{2}\\
&  \leqslant\frac{1}{\left(  n-\left\vert j\right\vert \right)  ^{4}
}E\left\vert \sum_{t=\left\vert j\right\vert +1}^{n}\psi_{t}\right\vert
^{2}\left\vert \sum_{t=\left\vert j\right\vert +1}^{n}\psi_{t-\left\vert
j\right\vert }\left(  v\right)  \right\vert ^{2}\\
&  \leqslant\frac{1}{\left(  n-\left\vert j\right\vert \right)  ^{4}}\left\{
E\left\vert
{\displaystyle\sum_{t=\left\vert j\right\vert +1}^{n}}
\psi_{t}\right\vert ^{4}E\left\vert
{\displaystyle\sum_{t=\left\vert j\right\vert +1}^{n}}
\psi_{t-\left\vert j\right\vert }\left(  v\right)  \right\vert ^{4}\right\}
^{1/2}%
\end{align*}
where we need two distinct\ conditions satisfied in order to prove that the
last term is uniformly bounded in $v\in\mathbb{R}.$ First, as seen in
Fokianos, $E\left\vert
{\displaystyle\sum_{t=\left\vert j\right\vert +1}^{n}}
\psi_{t-\left\vert j\right\vert }\left(  v\right)  \right\vert ^{4}\leqslant
C\left(  n-j\right)  ^{2}$ (Doukhan and Louhichi, 1999, Lemma 6)$.$
Additionally, it is required that the term $E\left\vert
{\displaystyle\sum_{t=\left\vert j\right\vert +1}^{n}}
\psi_{t}\right\vert ^{4}=E\left\vert
{\displaystyle\sum_{t=\left\vert j\right\vert +1}^{n}}
\left(  X_{t}-E\left(  X_{t}\right)  \right)  \right\vert ^{4}$is
bounded. We can write, additionally%
\begin{align*}
&  E\left\vert \widetilde{\sigma}_{j}\left(  v\right)  \right\vert ^{2}
  =E\left\vert \frac{1}{n-\left\vert j\right\vert }
{\displaystyle\sum_{t=\left\vert j\right\vert +1}^{n}}
\psi_{t}\psi_{t-\left\vert j\right\vert }\left(  v\right)  \right\vert ^{2}
  \leqslant\frac{C}{\left(  n-\left\vert j\right\vert \right)  }\\
\end{align*}
uniformly. \end{proof}

\begin{lemma}
\bigskip Suppose that $\left\{  X_{t},\text{ }t\geqslant1\right\}  $ satisfies
Assumptions 1 and 3(ii). For $\gamma>0$, denote by $D\left(  \gamma\right)  $
the region $D\left(  \gamma\right)  =\left\{  v:\gamma<\left\vert v\right\vert
<1/\gamma\right\}  .$ Then, under Assumption 4, for any fixed $\gamma>0,$
\[
\int_{D\left(  \gamma\right)  }\sum_{j=1}^{n-1}k^{2}\left(  j/p\right)
\left(  n-j\right)  \left\{  \left\vert \widehat{\sigma}_{j}\left(  v\right)
\right\vert ^{2}-\left\vert \widetilde{\sigma}_{j}\left(  v\right)
\right\vert ^{2}\right\}  d\mathcal{W=O}_{P}\left(  p/\sqrt{n}\right)
=o_{P}\left(  \sqrt{p}\right)
\]
as $p/n\rightarrow0$. The result of the Lemma is also true under independence.
\end{lemma}

\bigskip

\begin{proof}
[Proof of Lemma 3]
The proof of the lemma is exactly the same as in Fokianos and Pitsillou (2017)\end{proof}

\bigskip

\ \ \ \ \ 

\begin{proof}
[Proof of Proposition 1]Let us first define
\begin{align*}
\sigma_{j}\left(  v\right)   &  =cov\left(  X_{t},e^{ivX_{t-\left\vert
j\right\vert }}\right) \\
&  =E\left(  X_{t}e^{ivX_{t-j}}\right)  -E\left(  X_{t}\right)  E\left(
e^{ivX_{t-j}}\right)
\end{align*}
where $\phi\left(  v\right)  =E\left(  e^{ivX_{t-j}}\right)  .$ We have
\begin{align*}
MDD_{X}^{2}\left(  j\right)   &  =\int_{
\mathbb{R}
}\left\vert \sigma_{j}\left(  v\right)  \right\vert ^{2}d\mathcal{W}\left(
v\right) \\
&  =\left\vert \left\vert \sigma_{j}\left(  v\right)  \right\vert \right\vert
_{\mathcal{W}}^{2}%
\end{align*}
Let us also define the random variable%
\[
\widehat{\sigma}_{j}\left(  v\right)  =\frac{1}{n-\left\vert j\right\vert
}\sum_{t=\left\vert j\right\vert +1}^{n}X_{t}e^{ivX_{t-\left\vert j\right\vert
}}-\frac{1}{n-\left\vert j\right\vert }\sum_{t=\left\vert j\right\vert +1}%
^{n}X_{t}\frac{1}{n-\left\vert j\right\vert }\sum_{t=\left\vert j\right\vert
+1}^{n}e^{ivX_{t-\left\vert j\right\vert }}%
\]
and, after elementary transformations, by denoting%
\begin{align*}
\psi_{t}  &  =X_{t}-E\left(  X_{t}\right) \\
\psi_{t-j}  &  =e^{ivX_{t-\left\vert j\right\vert }}-\phi\left(  v\right)
\end{align*}
we can write
\[
\widehat{\sigma}_{j}\left(  v\right)  =\frac{1}{n-\left\vert j\right\vert
}\sum_{t=\left\vert j\right\vert +1}^{n}\psi_{t}\psi_{t-j}\left(  v\right)
-\frac{1}{n-\left\vert j\right\vert }\sum_{t=\left\vert j\right\vert +1}%
^{n}\psi_{t}\frac{1}{n-\left\vert j\right\vert }\sum_{t=\left\vert
j\right\vert +1}^{n}\psi_{t-j}\left(  v\right)
\]
where $\phi\left(  v\right)  =E\left(  e^{ivX_{t-\left\vert j\right\vert }%
}\right)  .$ Therefore, we can pose%
\[
\widehat{MDD}_{X}^{2}\left(  j\right)  =\left\vert \left\vert \widehat{\sigma
}_{j}\left(  v\right)  \right\vert \right\vert ^{2}=\int_{
\mathbb{R}
}\left\vert \widehat{\sigma}_{j}\left(  v\right)  \right\vert ^{2}%
d\mathcal{W}\left(  v\right)
\]
Let us now, for each $\gamma>0,$ describe the region
\[
D\left(  \gamma\right)  =\left\{  v:\gamma\leq\left\vert v\right\vert
\leq1/\gamma\right\}
\]
such that, by construction, $\gamma\in(0,1]$ and $\gamma\rightarrow
0\Longrightarrow$ $D\left(  \gamma\right)  =%
\mathbb{R}
.$ Define now the integral%
\[
\widehat{MDD}_{X;\gamma}^{2}\left(  j\right)  =\int_{D\left(  \gamma\right)
}\left\vert \widehat{\sigma}_{j}\left(  v\right)  \right\vert ^{2}%
d\mathcal{W}\left(  v\right)
\]
By the SLLN for $\alpha$-mixing random variables and the fact that
$\int_{D\left(  \gamma\right)  }d\mathcal{W<\infty}$, we obtain%
\[
\lim_{n\rightarrow\infty}\widehat{MDD}_{X;\gamma}^{2}\left(  j\right)
=MDD_{X;\gamma}^{2}\left(  j\right)
\]
almost surely. Clearly $MDD_{X;\gamma}^{2}\left(  j\right)  \rightarrow
MDD_{X}^{2}\left(  j\right)  $ as $\gamma\rightarrow0$.
We are therefore left to prove that almost surely%
\[
\lim\sup_{\gamma\rightarrow0}\lim\sup_{n\rightarrow\infty}\left\vert
\widehat{MDD}_{X;\gamma}^{2}\left(  j\right)  -\widehat{MDD}_{X}^{2}\left(
j\right)  \right\vert =0
\]
For each $\gamma>0$ we obtain that%
\begin{align*}
&  \left\vert \widehat{MDD}_{X;\gamma}^{2}\left(  j\right)  -\widehat{MDD}
_{X}^{2}\left(  j\right)  \right\vert \\
&  =\left\vert \int_{D\left(  \gamma\right)  }\left\vert \widehat{\sigma}
_{j}\left(  v\right)  \right\vert ^{2}d\mathcal{W}\left(  v\right)  -\int_{
\mathbb{R}
}\left\vert \widehat{\sigma}_{j}\left(  v\right)  \right\vert ^{2}
d\mathcal{W}\left(  v\right)  \right\vert \\
&  =\left\vert \int_{D\left(  \gamma\right)  }\left\vert \widehat{\sigma}
_{j}\left(  v\right)  \right\vert ^{2}d\mathcal{W}\left(  v\right)
-\int_{D\left(  \gamma\right)  }\left\vert \widehat{\sigma}_{j}\left(
v\right)  \right\vert ^{2}d\mathcal{W}\left(  v\right)  -\int_{D\left(
\gamma\right)  ^{c}}\left\vert \widehat{\sigma}_{j}\left(  v\right)
\right\vert ^{2}d\mathcal{W}\left(  v\right)  \right\vert \\
&  =\int_{D\left(  \gamma\right)  ^{c}}\left\vert \widehat{\sigma}_{j}\left(
v\right)  \right\vert ^{2}d\mathcal{W}\left(  v\right) \\
&  =\int_{\left\vert v\right\vert <\gamma}\left\vert \widehat{\sigma}
_{j}\left(  v\right)  \right\vert ^{2}d\mathcal{W}\left(  v\right)
+\int_{\left\vert v\right\vert >1/\gamma}\left\vert \widehat{\sigma}
_{j}\left(  v\right)  \right\vert ^{2}d\mathcal{W}\left(  v\right)
\end{align*}
and therefore%
\[
\left\vert \widehat{MDD}_{X;\gamma}^{2}\left(  j\right)  -\widehat{MDD}%
_{X}^{2}\left(  j\right)  \right\vert \leq\int_{\left\vert v\right\vert
<\gamma}\left\vert \widehat{\sigma}_{j}\left(  v\right)  \right\vert
^{2}d\mathcal{W}\left(  v\right)  +\int_{\left\vert v\right\vert >1/\gamma
}\left\vert \widehat{\sigma}_{j}\left(  v\right)  \right\vert ^{2}%
d\mathcal{W}\left(  v\right)
\] Now, for $z\in%
\mathbb{R}
$, define the function%
\[
H\left(  y\right)  =\int_{\left\vert z\right\vert <y}\frac{1-\cos\left(
z\right)  }{\left\vert z\right\vert ^{2}}dz
\]
such that it is bounded by $1/\pi^{2}$ and $\lim_{y\rightarrow0}H\left(
y\right)  =0.$ This can be shown as follows: for $d=1$ , $\alpha=1$ and $x=1,$
by using the result of Lemma 1:%
\begin{align*}
\int_{\left\vert z\right\vert <y}\frac{1-\cos\left(  z\right)  }{\left\vert
z\right\vert ^{2}}dz  &  \leq\int_{
\mathbb{R}
}\frac{1-\cos\left(  z\right)  }{\left\vert z\right\vert ^{2}}dz\\
&  =\pi
\end{align*}
being the integration of a nonnegative term for every $z\in%
\mathbb{R}
$. Because of the relation $\left\vert x+y\right\vert ^{2}\leq2\left\vert
x\right\vert ^{2}+2\left\vert y\right\vert ^{2}$ we have%
\begin{align*}
\left\vert \widehat{\sigma}_{j}\left(  v\right)  \right\vert ^{2}  &
=\left\vert \frac{1}{n-\left\vert j\right\vert }\sum_{t=\left\vert
j\right\vert +1}^{n}\psi_{t}\psi_{t-j}\left(  v\right)  -\frac{1}{n-\left\vert
j\right\vert }\sum_{t=\left\vert j\right\vert +1}^{n}\psi_{t}\frac
{1}{n-\left\vert j\right\vert }\sum_{t=\left\vert j\right\vert +1}^{n}
\psi_{t-j}\left(  v\right)  \right\vert ^{2}\\
&  \leq2\left\vert \frac{1}{n-\left\vert j\right\vert }\sum_{t=\left\vert
j\right\vert +1}^{n}\psi_{t}\psi_{t-j}\left(  v\right)  \right\vert
^{2}+2\left\vert \frac{1}{n-\left\vert j\right\vert }\sum_{t=\left\vert
j\right\vert +1}^{n}\psi_{t}\frac{1}{n-\left\vert j\right\vert }
\sum_{t=\left\vert j\right\vert +1}^{n}\psi_{t-j}\left(  v\right)  \right\vert
^{2}%
\end{align*}
For the first term, the Cauchy-Schwarz inequality for sums implies%
\[
2\left\vert \frac{1}{n-\left\vert j\right\vert }\sum_{t=\left\vert
j\right\vert +1}^{n}\psi_{t}\psi_{t-j}\left(  v\right)  \right\vert ^{2}%
\leq\frac{2}{\left(  n-\left\vert j\right\vert \right)  ^{2}}\sum
_{t=\left\vert j\right\vert +1}^{n}\left\vert \psi_{t}\right\vert ^{2}%
\sum_{t=\left\vert j\right\vert +1}^{n}\left\vert \psi_{t-j}\left(  v\right)
\right\vert ^{2}%
\]
and we also have%
\begin{align*}
2\left\vert \frac{1}{n-\left\vert j\right\vert }\sum_{t=\left\vert
j\right\vert +1}^{n}\psi_{t}\frac{1}{n-\left\vert j\right\vert }
\sum_{t=\left\vert j\right\vert +1}^{n}\psi_{t-j}\left(  v\right)  \right\vert
^{2}  &  \leq2\frac{1}{\left(  n-\left\vert j\right\vert \right)  ^{4}}
\sum_{t=\left\vert j\right\vert +1}^{n}\left\vert \psi_{t}\right\vert ^{2}
\sum_{t=\left\vert j\right\vert +1}^{n}\left\vert \psi_{t-j}\left(  v\right)
\right\vert ^{2}\\
&  \leq2\frac{1}{\left(  n-\left\vert j\right\vert \right)  ^{2}}
\sum_{t=\left\vert j\right\vert +1}^{n}\left\vert \psi_{t}\right\vert ^{2}
\sum_{t=\left\vert j\right\vert +1}^{n}\left\vert \psi_{t-j}\left(  v\right)
\right\vert ^{2}%
\end{align*}
and we finally get to the relation%
\[
\left\vert \widehat{\sigma}_{j}\left(  v\right)  \right\vert ^{2}\leq4\left\{
\frac{1}{\left(  n-\left\vert j\right\vert \right)  }\sum_{t=\left\vert
j\right\vert +1}^{n}\left\vert \psi_{t}\right\vert ^{2}\right\}  \left\{
\frac{1}{\left(  n-\left\vert j\right\vert \right)  }\sum_{t=\left\vert
j\right\vert +1}^{n}\left\vert \psi_{t-\left\vert j\right\vert }\left(
v\right)  \right\vert ^{2}\right\}
\]
By taking the integral%
\[
\int_{\left\vert v\right\vert <\gamma}\left\vert \widehat{\sigma}_{j}\left(
v\right)  \right\vert ^{2}d\mathcal{W}\left(  v\right)  \leq\left\{  \frac
{4}{\left(  n-\left\vert j\right\vert \right)  }\sum_{t=\left\vert
j\right\vert +1}^{n}\left\vert \psi_{t}\right\vert ^{2}\right\}  \left\{
\frac{1}{\left(  n-\left\vert j\right\vert \right)  }\sum^{n}\int_{\left\vert
v\right\vert <\gamma}\left\vert \psi_{t-\left\vert j\right\vert }\left(
v\right)  \right\vert ^{2}d\mathcal{W}\left(  v\right)  \right\}
\]
If we focus on the first factor on the left-hand side of the equation, we can
write%
\begin{align*}
\left\vert \psi_{t}\right\vert ^{2}  &  =\left\vert X_{t}-E\left(  X\right)
\right\vert ^{2}=\left\vert E_{X}(X_{t}-X)\right\vert ^{2}\\
&  \leq E_{X}\left\vert X_{t}-X\right\vert ^{2}\leq E_{X}\left\vert \left\vert
X_{t}\right\vert +\left\vert X\right\vert \right\vert ^{2}\\
&  \leq2E_{X}\left(  \left\vert X_{t}\right\vert ^{2}+\left\vert X\right\vert
^{2}\right) \\
&  =2\left(  \left\vert X_{t}\right\vert ^{2}+E\left\vert X_{1}\right\vert
^{2}\right)
\end{align*}
where the expectation $E_{X}$ is taken with respect to $X.$ Similarly, using Lemma 1 we show:%
\[
\int_{\left\vert v\right\vert <\gamma}\left\vert \psi_{t-\left\vert
j\right\vert }\left(  v\right)  \right\vert ^{2}d\mathcal{W}\left(  v\right)
\leq2E_{X}\left\vert X_{t-\left\vert j\right\vert }-X\right\vert H\left(
\left\vert X_{t-\left\vert j\right\vert }-X\right\vert \gamma\right)
\]
We have therefore%
\begin{align*}
&  \int_{\left\vert v\right\vert <\gamma}\left\vert \widehat{\sigma}
_{j}\left(  v\right)  \right\vert ^{2}d\mathcal{W}\left(  v\right) \\
&  \leq16\frac{1}{\left(  n-\left\vert j\right\vert \right)  }\sum
_{t=\left\vert j\right\vert +1}^{n}\left(  \left\vert X_{t}\right\vert
^{2}+E\left\vert X_{1}\right\vert ^{2}\right)  \frac{1}{\left(  n-\left\vert
j\right\vert \right)  }\sum_{t=\left\vert j\right\vert +1}^{n}E_{X}\left\vert
X_{t-\left\vert j\right\vert }-X\right\vert H\left(  \left\vert
X_{t-\left\vert j\right\vert }-X\right\vert \gamma\right)
\end{align*}
Because of Assumptions 1 and 2 and the ergodic theorem for $\alpha$-mixing
processes we obtain
\begin{align*}
\frac{1}{\left(  n-\left\vert j\right\vert \right)  }\sum_{t=\left\vert
j\right\vert +1}^{n}\left(  \left\vert X_{t}\right\vert ^{2}+E\left\vert
X_{1}\right\vert ^{2}\right)   &  \rightarrow E\left\vert X_{1}\right\vert
^{2}+E\left\vert X_{1}\right\vert ^{2}=2E\left\vert X_{1}\right\vert ^{2}\\
\frac{1}{\left(  n-\left\vert j\right\vert \right)  }\sum_{t=\left\vert
j\right\vert +1}^{n}E_{X}\left\vert X_{t-\left\vert j\right\vert
}-X\right\vert H\left(  \left\vert X_{t-\left\vert j\right\vert }-X\right\vert
\gamma\right)   &  \rightarrow E\left\vert X_{0}-X_{1}\right\vert H\left(
\left\vert X_{0}-X_{1}\right\vert \gamma\right)
\end{align*}
as $n\rightarrow\infty$ almost surely, so that%
\[
\lim\sup_{n\rightarrow\infty}\int_{\left\vert u\right\vert <\gamma}\left\vert
\widehat{\sigma}_{j}\left(  v\right)  \right\vert ^{2}d\mathcal{W}\left(
v\right)  \leq32E\left\vert X_{1}\right\vert ^{2}E\left\vert X_{0}%
-X_{1}\right\vert H\left(  \left\vert X_{0}-X_{1}\right\vert \gamma\right)
\]
Finally, by taking the limit for $\gamma\rightarrow0$%
\begin{align*}
&  \lim\sup_{\gamma\rightarrow0}\lim\sup_{n\rightarrow\infty}\int_{\left\vert
u\right\vert <\gamma}\left\vert \widehat{\sigma}_{j}\left(  v\right)
\right\vert ^{2}d\mathcal{W}\left(  v\right) \\
&  \leq32E\left\vert X_{1}\right\vert ^{2}\lim\sup_{\gamma\rightarrow
0}E\left\{  \left\vert X_{0}-X_{1}\right\vert H\left(  \left\vert X_{0}
-X_{1}\right\vert \gamma\right)  \right\}
\end{align*}
but being the term $\left\vert X_{0}-X_{1}\right\vert H\left(  \left\vert
X_{0}-X_{1}\right\vert \gamma\right)  $ 
bounded by assumption and being

 $\lim\sup_{\gamma\rightarrow0}\left[  \left\vert X_{0}-X_{1}\right\vert
H\left(  \left\vert X_{0}-X_{1}\right\vert \gamma\right)  \right]  =0,$ by
Lebesgue's dominated theorem we obtain that%
\begin{align*}
&  32E\left\vert X_{1}\right\vert ^{2}\lim\sup_{\gamma\rightarrow0}E\left\{
\left\vert X_{0}-X_{1}\right\vert H\left(  \left\vert X_{0}-X_{1}\right\vert
\gamma\right)  \right\} \\
&  =32E\left\vert X_{1}\right\vert ^{2}E\left\{  \lim\sup_{\gamma\rightarrow
0}\left[  \left\vert X_{0}-X_{1}\right\vert H\left(  \left\vert X_{0}
-X_{1}\right\vert \gamma\right)  \right]  \right\} \\
&  =0
\end{align*}
and therefore, we are able to show
\[
\lim\sup_{\gamma\rightarrow0}\lim\sup_{n\rightarrow\infty}\int_{\left\vert
u\right\vert <\gamma}\left\vert \widehat{\sigma}_{j}\left(  v\right)
\right\vert ^{2}d\mathcal{W}\left(  v\right)  \leq0
\]
Let us consider now the second term in the integral%
\[
\int_{\left\vert v\right\vert >1/\gamma}\left\vert \widehat{\sigma}_{j}\left(
v\right)  \right\vert ^{2}d\mathcal{W}\left(  v\right)  \leq\left\{  \frac
{4}{\left(  n-\left\vert j\right\vert \right)  }\sum_{t=\left\vert
j\right\vert +1}^{n}\left\vert \psi_{t}\right\vert ^{2}\right\}  \left\{
\frac{1}{\left(  n-\left\vert j\right\vert \right)  }\sum_{t=\left\vert
j\right\vert +1}^{n}\int_{\left\vert v\right\vert >1/\gamma}\left\vert
\psi_{t-\left\vert j\right\vert }\left(  v\right)  \right\vert ^{2}%
d\mathcal{W}\left(  v\right)  \right\}
\]
where it is implied that $\left\vert \psi_{t-\left\vert j\right\vert }\left(
v\right)  \right\vert ^{2}\leq4$ and $1/\left(  n-\left\vert j\right\vert
\right)  \sum_{t=\left\vert j\right\vert +1}^{n}\left\vert \psi_{t-\left\vert
j\right\vert }\left(  v\right)  \right\vert ^{2}\leq4.$ 

\medskip

Therefore we have:%
\begin{align*}
\int_{\left\vert v\right\vert >1/\gamma}\left\vert \widehat{\sigma}_{j}\left(
v\right)  \right\vert ^{2}d\mathcal{W}\left(  v\right)   &  \leq\left\{
\frac{4}{\left(  n-\left\vert j\right\vert \right)  }\sum_{t=\left\vert
j\right\vert +1}^{n}\left\vert \psi_{t}\right\vert ^{2}\right\}  \left\{
\frac{1}{\left(  n-\left\vert j\right\vert \right)  }\sum_{t=\left\vert
j\right\vert +1}^{n}\int_{\left\vert v\right\vert >1/\gamma}\left\vert
\psi_{t-\left\vert j\right\vert }\left(  v\right)  \right\vert ^{2}
d\mathcal{W}\left(  v\right)  \right\} \\
&  \leq\left\{  \frac{16}{\left(  n-\left\vert j\right\vert \right)  }
\sum_{t=\left\vert j\right\vert +1}^{n}\left\vert \psi_{t}\right\vert
^{2}\right\}  \left\{  \int_{\left\vert v\right\vert >1/\gamma}d\mathcal{W}
\left(  v\right)  \right\} \\
&  =\left\{  \frac{16}{\left(  n-\left\vert j\right\vert \right)  }
\sum_{t=\left\vert j\right\vert +1}^{n}\left\vert \psi_{t}\right\vert
^{2}\right\}  \left\{  \int_{\left\vert v\right\vert >1/\gamma}\frac{1}
{\pi\left\vert v\right\vert ^{2}}dv\right\}
\end{align*}
In order to solve the integral term in the second term of the multiplication we can write%
\[
\int_{\left\vert v\right\vert >1/\gamma}\frac{1}{\pi\left\vert v\right\vert
^{2}}dv=\int_{\left\vert v\right\vert >1/\gamma}\frac{1-\cos\left(  v\cdot
x\right)  }{\pi\left\vert v\right\vert ^{2}}dv
\]
for $x=1/v$. Thanks to Lemma 1 we can use the
relationship%
\[
\int_{\left\vert v\right\vert >1/\gamma}\frac{1-\cos\left(  v\cdot x\right)
}{\pi\left\vert v\right\vert ^{2}}dv=\left\vert x\right\vert =\left\vert
1/v\right\vert
\]
and by the fact that $\left\vert 1/v\right\vert \leq\gamma$ over the region of
integration, we can finally write%
\[
\int_{\left\vert v\right\vert >1/\gamma}\frac{1}{\pi\left\vert v\right\vert
^{2}}dv\leq\gamma
\]
Accordingly%
\begin{align*}
\int_{\left\vert v\right\vert >1/\gamma}\left\vert \widehat{\sigma}_{j}\left(
v\right)  \right\vert ^{2}d\mathcal{W}\left(  v\right)   &  \leq\frac
{16\gamma}{\left(  n-\left\vert j\right\vert \right)  }\sum_{t=\left\vert
j\right\vert +1}^{n}\left\vert \psi_{t}\right\vert ^{2}\\
&  \leq\frac{16\gamma}{\left(  n-\left\vert j\right\vert \right)  }
\sum_{t=\left\vert j\right\vert +1}^{n}2\left(  \left\vert X_{t}\right\vert
^{2}+E\left\vert X_{1}\right\vert ^{2}\right)
\end{align*}
Then, almost surely%
\[
\lim\sup_{\gamma\rightarrow0}\lim\sup_{n\rightarrow\infty}\int_{\left\vert
v\right\vert >1/\gamma}\left\vert \widehat{\sigma}_{j}\left(  v\right)
\right\vert ^{2}d\mathcal{W}\left(  v\right)  =0
\]and this is the end of our proof\end{proof}

\bigskip

\begin{proof}
[Proof of Theorem 2] The result is obtained following Theorem 1 in Fokianos and Pitsillou (2017) and Hong (1999). In particular, Hong's proofs are general and similarly valid in our setting.
\end{proof}

\bigskip

\begin{proof}
[Proof of Theorem 3]The result is obtained following Theorem 2 in Fokianos and Pitsillou (2017) and Hong (1999). In particular, Hong's proofs are general and similarly valid in our setting.
\end{proof}

\bigskip

\end{document}